\documentclass[12pt,preprint]{aastex}
\usepackage{rotating}

\sfcode`A=1000 \sfcode`B=1000 \sfcode`C=1000 \sfcode`D=1000
\sfcode`E=1000 \sfcode`F=1000 \sfcode`G=1000 \sfcode`H=1000
\sfcode`I=1000 \sfcode`J=1000 \sfcode`K=1000 \sfcode`L=1000
\sfcode`M=1000 \sfcode`N=1000 \sfcode`O=1000 \sfcode`P=1000
\sfcode`Q=1000 \sfcode`R=1000 \sfcode`S=1000 \sfcode`T=1000
\sfcode`U=1000 \sfcode`V=1000 \sfcode`W=1000 \sfcode`X=1000
\sfcode`Y=1000 \sfcode`Z=1000

\newcommand{\Msun}{\mbox{$M_\sun$}}
\newcommand{\Lsun}{\mbox{$L_\sun$}}
\newcommand{\0}{\phantom{0}}  
\newcommand{\no}{\nodata}

\newcommand{\AKARI}{{\it AKARI}}
\newcommand{\s}{{\it Spitzer}}
\newcommand{\iras}{{\it IRAS}}
\newcommand{\chandra}{{\it Chandra}}
\newcommand{\hii}{{\sc h~ii}}

\slugcomment{v2 \today}
\shorttitle{Star-forming Galaxies at $z\simeq1.8$}
\shortauthors{Huang et al.}
\received{}
\begin{document}
\title{IRS Spectroscopy and Multi-wavelength Study of Luminous
Star-forming Galaxies at $z \simeq 1.9$}
\author{J.-S.~Huang,\altaffilmark{1}
S.M.~Faber,\altaffilmark{2}
E.~Daddi,\altaffilmark{3}
E.~S.~Laird,\altaffilmark{4}
K.~Lai,\altaffilmark{1}
A.~Omont,\altaffilmark{5}
Y.~Wu,\altaffilmark{6}
J.~D.~Younger\altaffilmark{1}
K.~Bundy,\altaffilmark{7}
A.~Cattaneo, \altaffilmark{8}
S.~C.~Chapman,\altaffilmark{9}
C.J.~Conselice,\altaffilmark{10}
M.~Dickinson,\altaffilmark{11}
E.~Egami,\altaffilmark{12}
G.~G.~Fazio,\altaffilmark{1}
M. Im, \altaffilmark{13}
D.~Koo,\altaffilmark{2}
E.~Le Floc'h,\altaffilmark{14}
C.~Papovich,\altaffilmark{12}
D.~Rigopoulou,\altaffilmark{15}
I.~Smail,\altaffilmark{16}
M.~Song, \altaffilmark{13}
P.~P.~Van de Werf,\altaffilmark{17}
T.~M.~A.~Webb,\altaffilmark{18}
C.~N.~A.~Willmer,\altaffilmark{12}
S.~P.~Willner,\altaffilmark{1}
\&
L.~Yan,\altaffilmark{6} 
}

\altaffiltext{1}{Harvard-Smithsonian Center for Astrophysics, 60 Garden Street,
Cambridge, MA 02138}

\altaffiltext{2}{University of California Observatories/Lick
Observatory, University of California, Santa Cruz, CA 95064} 

\altaffiltext{3}{Laboratoire AIM, CEA/DSM-CNRS-Universit\'e Paris Diderot,
DAPNIA/Service d'Astrophysique, CEA Saclay, Orme des Merisiers, 91191
Gif-sur-Yvette Cedex, France}

\altaffiltext{4}{Astrophysics Group, Imperial College London,
Blackett Laboratory, Prince Consort Road, London SW7 2AZ}

\altaffiltext{5}{Institut d'Astrophysique de Paris-CNRS, 98bis
Boulevard Arago, F-75014 Paris, France}

\altaffiltext{6}{IPAC,
California Institute of Technology, 1200 E. California, Pasadena, CA 91125}

\altaffiltext{7}{Reinhardt Fellow, Department of Astronomy and Astrophysics, University of Toronto, Toronto, ON M5S 3H8, Canada}

\altaffiltext{8}{Astrophysikalisches Institut Potsdam, an der Sternwarte 16, 14482 Potsdam, Germany}

\altaffiltext{9}{Institute of Astronomy, Cambridge, CB3 0HA, UK.;
University of Victoria, Victoria, BC V8P 1A1 Canada}

\altaffiltext{10}{University of Nottingham, School of Physics \& Astronomy, Nottingham NG7 2RD}

\altaffiltext{11}{NOAO, 950 North Cherry Avenue,  Tucson, AZ 85719} 

\altaffiltext{12}{Steward Observatory, University of Arizona, 933
North Cherry Avenue, Tucson, AZ 85721}

\altaffiltext{13}{Department of Physics and Astronomy, FPRD, Seoul National University, Seoul 151-747, Korea}

\altaffiltext{14}{IfA, University of Hawaii, Honolulu, HI 96822}

\altaffiltext{15}{Department of Astrophysics, Oxford University,
Keble Road, Oxford, OX1 3RH, UK}

\altaffiltext{16}{Institute for Computational Cosmology, Durham
University, Durham, UK}

\altaffiltext{17}{Leiden Observatory, Leiden University, P.O. Box
9513, NL-2300 RA Leiden, Netherlands}

\altaffiltext{18}{Department of Physics, McGill University, Montr?al,
QC, Canada}

\begin{abstract}


  We analyze a sample of galaxies chosen to have $F_{24\micron} > 0.5
  mJy$ and satisfy a certain IRAC color criterion.  IRS spectra
  yield redshifts, spectral types, and PAH luminosities, to which we
  add broadband photometry from optical through IRAC wavelengths, MIPS
  from 24-160 \micron, 1.1 millimeter, and radio at 1.4 GHz.
  Stellar population modeling and IRS spectra together demonstrate
  that the double criteria used to select this sample have efficiently
  isolated massive star-forming galaxies at $z\sim 1.9$.  This is the
  first starburst-dominated ULIRG sample at high redshift with total
  infrared luminosity measured directly from FIR and millimeter
  photometry, and as such gives us the first accurate view of
  broadband SEDs for starburst galaxies at extremely high luminosity
  and at all wavelengths.  Similar broadband data are assembled for
  three other galaxy samples -- local starburst galaxies, local
  AGN/ULIRGS, and a second 24\micron-luminous $z\sim 2$ sample
  dominated by AGN.  $L_{PAH}/L_{IR}$ for the new $z\sim 2$ starburst
  sample is the highest ever seen, some three times higher than in
  local starbursts, whereas in AGNs this ratio is depressed below the
  starburst trend, often severely.  Several pieces of evidence imply
  that AGNs exist in this starburst dominated sample,   except two of which
  even host very strong AGN, while they still have very strong PAH
  emission.  The ACS images show most objects have very extended
  morphologies in the rest-frame UV band, thus extended distribution
  of PAH molecules. Such an extended distribution prevents further destruction
  PAH molecules by central AGNs.  We conclude that
  objects in this sample are ULIRGs powered mainly by starburst; 
  and the total infrared luminosity
  density contributed by this type of objects is $0.9-2.6\times 10^7
  L_{\odot}/Mpc^3$.


\end{abstract}
\keywords{cosmology: observations --- galaxies: dust emission  ---
  galaxies: mid-infrared} 

\section{INTRODUCTION}
  
Formation of massive galaxies provides a critical test of theories of
galaxy formation and evolution.  Before modern deep observations, the
most massive galaxies known were local elliptical galaxies with no
ongoing star formation.  The classical model for these objects
\citep[e.g.,][]{eggen1962} was monolithic formation at high redshifts,
followed by passive evolution.  A more recent galaxy formation theory
in the Cold Dark Matter (CDM) paradigm \citep{blumenthal1984,
  white1991,cole2000} predicts quite the opposite scenario: a
galaxy-galaxy merging-tree model. In this scenario, small galaxies
formed early in cosmic time, and massive galaxies were assembled later
at much lower redshifts by a series of mergers. Observations of local
Ultra-Luminous InfraRed Galaxies (ULIRGs, $L_{IR} >
10^{12}$~\Lsun)\footnote{$L_{IR}\equiv\int_{8~\micron}^{1000~\micron}
  L_{\nu} d\nu$ \citep{sanders1996}} detected by \iras\ are consistent
with the merger theory. Most local ULIRGs have disturbed morphologies,
consistent with being merging systems \citep{kim2002}. ULIRGs in later
stages of merging have $r^{-1/4}$ light profiles \citep{james1999,
  rothberg2004}.  \citet{genzel2001} and \citet{tacconi2002} measured
local ULIRG dynamical masses and found an average of
$10^{11}$~\Msun. These features are consistent with numerical
simulation studies of galaxy mergers, indicating that local ULIRGs are
merging systems transforming gas-rich galaxies into $L_*$ elliptical
galaxies \citep{kormendy1992, mihos1996, barnes1996, tacconi2002}.

The story is different at $z\ga2$.  Deep near-infrared surveys
\citep{franx2003, glazebrook2004, mccarthy2004, daddi2005, labbe2005}
have identified apparently luminous passive galaxies already in place
at $z \sim2$, implying that they formed at even higher redshifts.  The
existence of galaxies with $M_*>10^{11}$~\Msun\ at high redshifts may
challenge the merger theory of forming such objects at lower
redshifts.  However, \citet{cattaneo2008} used a semi-analytic model
to show that significant numbers of $M_*>10^{11} M_{\odot}$ galaxies
were in places by $z\sim 2$ but many also formed at lower redshifts,
that is, there was a whole "downsizing" trend for massive galaxies to
form their stars early, but it is merely statistical not
absolute. Possibly consistent with this is the fact that the
contribution of LIRGs and ULIRGs to the total IR luminosity density is
more than 70\% at $z=1$ \citep{lefloch2005} compared to a negligible
percentage locally \citep{sanders1996, huang2007b}.  Moreover, the
redshift surveys for Sub-Millimeter Galaxies (SMGs) by
\citet{chapman2003, chapman2006} reveal a rapidly evolving ULIRG
population at $1.7<z<2.8$.  Such strong evolution is also seen in
ULIRGs selected with {\it BzK} color and MIPS 24~\micron\ flux at
$z\sim 2$ with their number density apparently 3 orders of magnitude
higher than the local number density.  Thus local ULIRGs would well be
the tail end of earlier intense activity.

The Spitzer MIPS 24~\micron\ band has been very effective in probing
infrared emission from galaxies at redshifts up to $z\sim 3$
\citep{huang2005,webb2006,papovich2006,rigopoulou2006, daddi2005,
  daddi2007a, daddi2007b}. \citet{papovich2006}, \citet{webb2006},
\citet{daddi2007a, daddi2007b}, and \citet{dey2008} argued that
24~\micron\ emission from galaxies at $2<z<3$ is powered by both
active galactic nuclei (AGN) and star formation.  Spectroscopic
observations of a few 24~\micron\ luminous SMGs and Lyman break
galaxies (LBGs) at $1<z<3$ \citep{lutz2005, huang2007a, lefloch2007,
  valiante2007, pope08} with the Infrared Spectrograph (IRS) on
Spitzer support this view, showing both strong continua and emission
features of Polycyclic Aromatic Hydrocarbons (PAH) in the rest-frame
$6 < \lambda < 10$~\micron.  Systematic infrared
spectroscopic surveys of 24~\micron\ luminous but optically faint
sources \citep{houck2005, yan2005} reveal a dusty, $z \sim 2$ AGN
population not detected in optical surveys. Most of these AGNs are
ULIRGs with power-law spectral energy distributions (SEDs) in the
mid-infrared and deep silicate absorption at 9.7~\micron\
\citep{sajina2007a}.  \citet{weedman2006} observed a sample of X-ray
AGN with similar properties though generally less silicate absorption.
Optically-faint radio sources are a mix of AGN and starbursts
\citep{Weedman2006a} but are predominantly AGN.  In general,
optically-faint objects have weak infrared spectral emission features,
and most objects are likely to be AGN \citep{Weedman2006c}. However,
not all 24~\micron\ luminous objects at $z\sim 2-3$ are
AGN-dominated. For example, \citet{weedman2006} and \citet{farrah2008}
also identified samples with an apparent 1.6~\micron\ stellar peak in
the IRAC 4.5 or 5.8~\micron\ bands.  Both samples show a very narrow
redshift distribution due to a selection of the MIPS 24~\micron\ band
toward strong 7.7\micron\ PAH emission at $z\sim 1.9$.  IRS
spectroscopy of 24~\micron-luminous SMGs
\citep{lutz2005,Menendez2007,valiante2007, pope08,Menendez2008} shows
similar spectral features, namely strong PAH emission in objects with
a 1.6\micron\ stellar emission bump \citep{weedman2006,farrah2008},
indicating intensive star formation in both types of objects.

This paper presents an IRS spectroscopic and multi-wavelength study of
a ULIRG sample at $z\sim 2$.  The sample comes from the All-wavelength
Extended Groth-strip International Survey \citep[AEGIS,][]{davis2007},
which consists of deep surveys ranging from X-ray to radio
wavelengths.  Selection of our sample catches a starburst-dominated
phase of ULIRG with $L_{IR}>5\times10^{12}$L$_{\odot}$, which is very
rare among local ULIRGs. In this paper, we will study their properties
including star formation, stellar masses, AGN fractions, and
contribution to the universe's star formation history.  \S2 describes
the sample selection.  The IRS spectroscopic results are presented in
\S3, and \S4 contains an analysis of stellar populations, star
formation rate, and AGN fraction.  \S5 summarizes our results.  All
magnitudes are in the AB magnitude system unless stated otherwise, and
notation such as ``[3.6]'' means the AB magnitude at wavelength
3.6~\micron. The adopted cosmology parameters are
$H_0=70$~km~s$^{-1}$~Mpc$^{-1}$, $\Omega_M = 0.3$, $\Omega_\Lambda =
0.7$.
 
\section{SAMPLE SELECTION}

We wish to study the multi-wavelength properties of star-forming
galaxies at $z\sim 2$.  There are many ways of using optical and NIR
colors to select such a sample. The samples with the best
spectroscopic confirmation are the $UgR$ color-selected BM/BX sources
\citep{steidel2004}, which have estimated typical stellar masses of
about $10^9 \sim 10^{10}$~\Msun\ \citep{shapley2005, reddy2006,
  reddy2007}.  The average 24~\micron\ flux density for these sources
is $42\pm6$~$\mu$Jy \citep{reddy2006}, which suggests modest rest
frame mid-IR luminosities, consistent with LIRGs
($L_{IR}<10^{12}$~\Lsun).  A different sample, based on near-infrared
color selection, is the Distant Red Galaxies (DRG, $(J-K)_{vega}>2.3$)
\citep{franx2003, labbe2005}.  These galaxies are redder and dustier
than the UV-selected BM/BX sources and are believed to be more massive
than $10^{11}$\Msun\ \citep{labbe2005, papovich2006}. Dusty DRGs have
estimated total infrared luminosity in the range $10^{11}<L_{IR}<
10^{12} \Lsun$ \citep{webb2006,papovich2006}. A third sample
\citep{daddi2005} uses {\it BzK} colors to select galaxies at $z\sim
2$; massive {\it BzK} galaxies are mid-IR luminous.  \citet{reddy2005}
compared BM/BX, DRGs, and {\it BzK} galaxies and found that {\it BzK}
galaxies include most DRGs and BM/BX galaxies.  This comparison is
nicely shown in Fig.~9 of \citet{reddy2005}.

\begin{figure}
\plotone{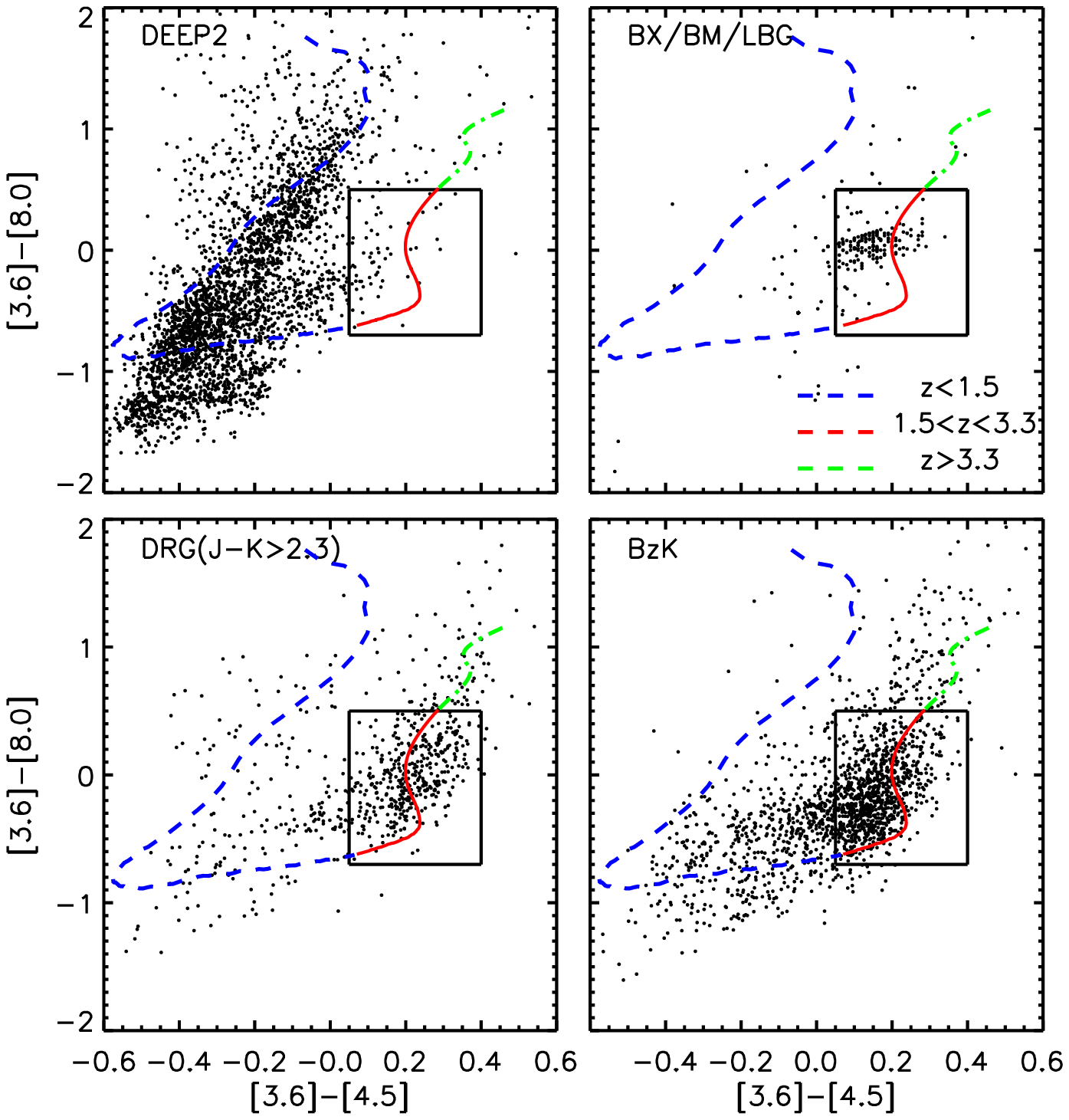}
\caption{IRAC color-color diagram for several samples. The upper left
panel is for the entire AEGIS spectroscopic redshift sample with
$0<z<1.5$; the upper right panel is for the combined BM/BX and LBG
samples with confirmed spectroscopic redshifts \citep{steidel2004, reddy2007}; the lower left
panel is for the DRGs \citep{franx2003}; and the lower right panel is
for BzK galaxies \citep{daddi2005}.  The boxes in each panel denote
the IRAC color selection for the present sample.  The track for the M82 template
is also plotted in each panel. The rest-frame
UV-selected BM/BX and LBG galaxies have generally faint IRAC flux
densities and thus larger photometric uncertainties \citep{huang2005,
rigopoulou2006}, increasing the apparent scatter in the upper right
panel. 
\label{f:cc2}}
\end{figure}

An independent way to select galaxies at $z>1.5$ is to use IRAC
colors.  In $1.5<z<3.0$, the four IRAC bands probe the rest-frame NIR
bands where galaxy SEDs have similar shape, thus the IRAC colors are
very robust in determining redshift in this range\citep{huang2004,
  papovich2008}.  The 1.6~\micron\ stellar emission bump can be used
to separate galaxies at $z=1.5$.  At $z<1.5$, the IRAC 3.6 and
4.5~\micron\ bands sample galaxy SED from the Jeans tail of cold
stars, thus the $[3.6]-[4.5]<0$.  At $z \ga 1.5$, the 1.6~\micron\
stellar emission bump begins to move into the IRAC 4.5~\micron\ band,
making the $[3.6]-[4.5]>0$.  The color criteria is set based on the
M82 SED model \citep{huang2004}
\begin{equation}
       0.05<[3.6]-[4.5]<0.4,~~~and
\end{equation}
\begin{equation}
       -0.7<[3.6]-[8.0]<0.5~~~~~~~~~
\end{equation}
Both color ranges are corresponding to redshift range of $1.5 \la z
\la 3.3$.  The red color cut in both equations rejects power-law AGNs
and star forming galaxies at $z>3.3$ whose 7.7\micron\ PAH shifts out
of the IRS wavelength range. The selection is based on the
rest-frame NIR colors, and is thus less affected by dust extinction,
stellar ages, and metallicities.  
Figure~\ref{f:cc2} compares IRAC two-color plots for the DEEP2
galaxies with $z<1.5$ to LBGs, DRGs, and {\it BzK} galaxies. 
The color criteria of equations 1 and 2 include most galaxies
in the $1.5 \la z \la 3.0$ range.
Although 8~\micron\ detection is required for this selection method,
selecting at this wavelength has some additional advantages.  Chosen
at roughly 2--3~\micron\ rest-frame, the sample selection is immune to
dust reddening and is roughly equivalent to a stellar-mass-selected
sample. It is thus ideal for studying luminous, potentially massive
galaxies \citep{huang2004, huang2005, conselice2007}.

\begin{figure}
\plotone{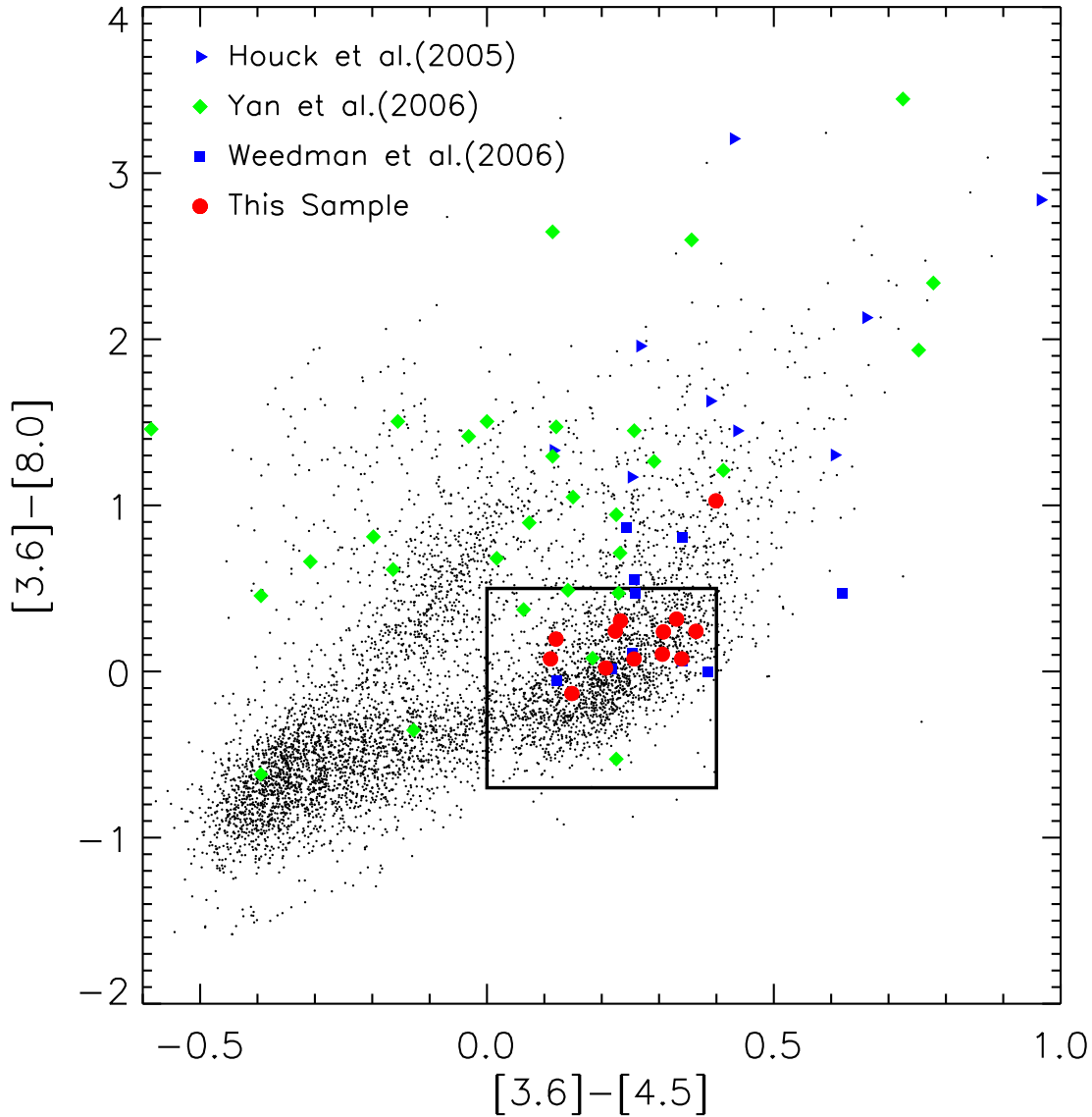}
\caption{IRAC color-color diagram for EGS galaxies with 
$F(24~\micron)>80$~$\mu$Jy. Small dots show all such galaxies; large
red dots show galaxies in the current IRS spectroscopic sample, which
also requires $F(24~\micron)>500$~$\mu$Jy.
The black box shows the IRAC color 
criteria, which should select objects at $z>1.5$.  The one red dot outside the
selection box is the serendipitous source EGS24a.
Objects from other IRS spectroscopic samples (Table~1) at $z\sim 2$
are  plotted for comparison: blue triangles and green diamonds denote ``optically
invisible'' sources
\citep{houck2005, yan2005}.
Blue squares denote the luminous starbursts of \citet{weedman2006}.
\label{f:cc}}
\end{figure}

The specific IRS targets for this program were selected from a
24~\micron\ sample \citep{papovich2004} in the EGS region. These
sources show a clump at the predicted colors for $1.5<z<3$ in
Figure~\ref{f:cc}, but redshifts were not known in advance.
Individual targets were selected to have IRAC colors satisfying Eqs. 1
and 2 and also $F_{24\micron} > 0.5$~mJy.  In addition, each candidate
was visually examined in the deep Subaru R-band image
\citep{ashby2008} to avoid confused or blended targets. With these
criteria, 12 targets were selected in the 2\arcdeg$\times$10\arcmin\
EGS region for IRS observation.  Table~1 lists the sample galaxies.
In this redshift range, most sources will be either ULIRGs with total
infrared luminosity $L_{IR}>10^{12}$~\Lsun\ 
\footnote{ The total infrared luminosites ($L_{IR}$) for our sample are 
calculated with MIPS
  24, 70, 160~\micron\, and 1.1mm flux densities and Chary-Elbaz
  \citep{chary2001, daddi2007a, daddi2007b} SED models.  Details are
  given in \S\ref{s:lir}.} or AGNs with high mid-IR
luminosities. For convenience, the galaxy nicknames used
in the \s\ database are used in this paper, but these do not follow
proper naming conventions and should not be used as sole object
identifiers. Proper names are also given in Table 1.

Most of the previous IRS surveys of IR luminous sources at $z\sim 2$
have used rather different selection criteria. Table~2 summarizes the
sample criteria for various other IRS surveys. \citet{houck2005} and
\citet{yan2005} used extreme optical-to-24~\micron\ color to select
dusty objects. Objects in these samples have much redder [3.6]-[8.0]
IRAC colors than the majority of 24~\micron\ sources (Fig.~\ref{f:cc})
and are mostly AGNs as shown by their strong power-law continua, but
weak or absent PAH emission features
\citep{houck2005,yan2005}. \citet{weedman2006} selected AGN using
similar criteria.  They also selected a separate starburst-dominated
sample at $z\sim 2$ based on the stellar 1.6~\micron\ emission bump.
The exact criterion required the peak flux density to be at either
4.5~\micron\ or 5.8~\micron, thus rejects low-redshift galaxies and
AGN with strong power-law SEDs . The resulting sample is very similar
to ours though overall a bit redder (Fig.~\ref{f:cc}).  All objects in
the \citeauthor{weedman2006} starburst sample show strong PAH emission
features.


\section{IRS Spectroscopy}
\subsection{Observations and Data Reduction}

IRS observations of this sample are part of the GTO program for the
\s/IRAC instrument team (program ID:\ 30327).  Objects were observed
only with the IRS Long-Slit Low-Resolution first order (LL1) mode,
giving wavelength coverage $20<\lambda <38$~\micron\ with spectral
resolution $60 \la \lambda/\Delta\lambda \la 120$.  The wavelength
coverage corresponds to $6\la \lambda \la13$~\micron\ in the
rest-frame for galaxies at $z \approx 2$. This wavelength range
includes strong PAH emission features at 7.7, 8.6, and 11.3~\micron\
and silicate absorption from 8 to 13~\micron\ (peaking near
9.7~\micron).  Detecting these features permits redshift measurement
and study of dust properties.  Total exposure time for each target was
based on its 24~\micron\ flux density.  Mapping mode
\citep{teplitz2007} was used to place each object at 6 positions
spaced 24\arcsec\ apart along the 168\arcsec\ IRS slit.  This not only
gives more uniform spectra for the target objects, rejecting cosmic
rays and bad pixels, but also increases sky coverage for possible
serendipitous objects around each target.  Table~1 gives the target
list and other parameters for the observations.  All data were
processed with the Spitzer Science Center pipeline, version
13.0. Extraction of source spectra was done with both the SMART
analysis package \citep{higdon2004} and our customized software. Lack
of IRS coverage at $\lambda \la 20$~\micron\ for this sample is
compensated with deep \AKARI\ 15~\micron\ imaging \citep{im2008}.  All
objects except two outside the \AKARI\ area are detected at
15~\micron\, providing measurement of the continua at rest-frame $\sim
6$~\micron\ for galaxies at $z \approx 2$.

\begin{figure}
\plotone{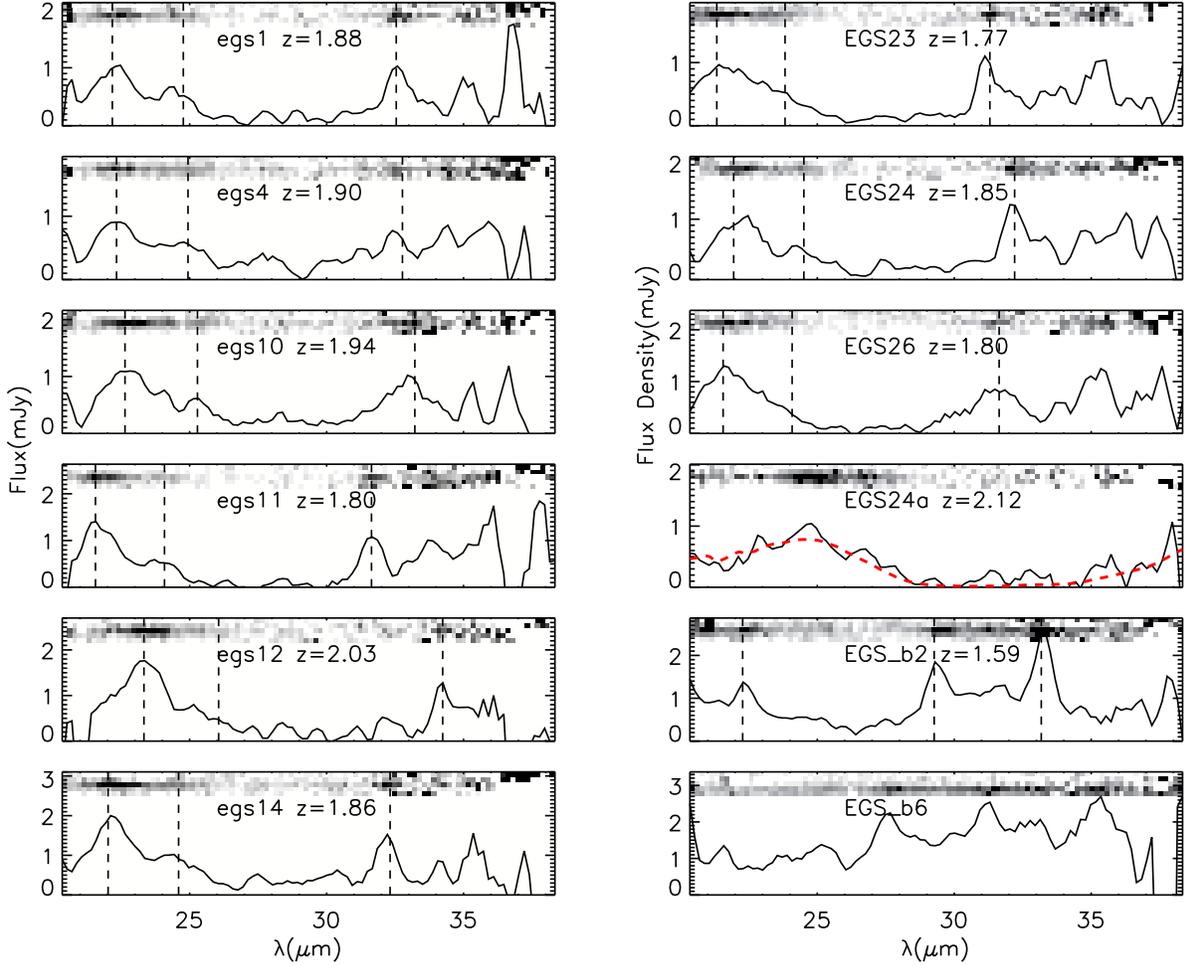}
\caption{Observed IRS spectra. The vertical scale is linear but
different for each panel.  The gray-scale images are the
two-dimensional IRS spectral images after wavelength and position
calibration.  Each image shows 5 pixels or 25\farcs5 along the slit.
The dashed lines indicate the central wavelengths of the
PAH emission features,  rest-frame 7.7, 8.6, and 11.3~\micron\ left to right. 
EGS24a, the serendipitous object in the
slit of EGS24, shows a power-law SED with
strong silicate absorption. Cross-correlation of the template (red dashed line)
of the local ULIRG IRAS F08572+3915 to the spectrum of EGS24a gives $z=2.12$.
 For EGS\_b2 at $z=1.59$, the 7.7~\micron\ feature is off scale to the
left, and the peak observed at 33.5~\micron\ is the [\ion{Ne}{2}]
emission line at rest wavelength 12.81~\micron. EGS\_b6 is a confused case with combining spectra
of two galaxies at z=1.02 and z=2.0. 
\label{f:spec}}
\end{figure}

\begin{figure}
\begin{center}
\plotone{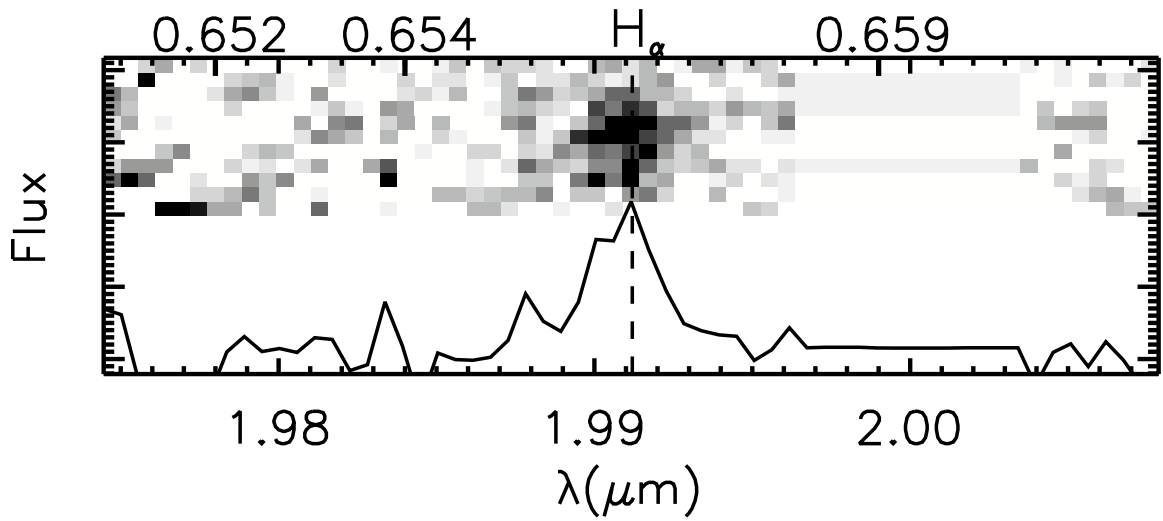}
\end{center}
\caption{Near infrared spectrum of EGS12 taken with the MOIRC
spectragraph on Subaru. There is nothing detected except one emission
line at 1.992~\micron.  We identify this line as H${\alpha}$ at
$z=2.033$, and corresponding rest-frame wavelengths are marked above the plot.  The
redshift is consistent with $z=2.03$ derived from the PAH features in
the IRS spectrum (Fig.~\ref{f:spec}).
\label{f:nir_spec}}
\end{figure}

Figure~\ref{f:spec} presents the IRS spectra.  PAH emission features
at 7.7 and 11.3~\micron\ and silicate absorption peaking at
9.7~\micron\ are detected from 10 sources, indicating a narrow
redshift range of $1.6<z<2.1$.  The PAH emission features at 7.7 and
11.3~\micron\ show pronounced variations in their profiles and peak
wavelengths. Both 7.7 and 11.3~\micron\ PAH features have at least two
components \citep{Peeters2002}. For example, the 7.7~\micron\ PAH
feature has a blue component at 7.6~\micron\ and a red component at
wavelength longwards of 7.7~\micron.  Thus different types of PAH
spectral templates potentially yield different redshift measurements.
To check this, we use two local MIR spectral templates with different
PAH profiles, an average local starburst spectrum and an average local
ULIRG spectrum to determine redshifts.  Both templates yield very
similar redshifts (Table~3). The starburst template fits all objects
better with a typical 2\% redshift uncertainty.  EGS\_b2 is identified
at $z=1.59$ with PAH emission features at 8.6 and 11.3~\micron\ and
the [\ion{Ne}{2}] emission line at 12.81~\micron.  Redshift $z=2.03$
for EGS12 is confirmed by detecting $H{\alpha}$ at 1.992~\micron\
(Figure~\ref{f:nir_spec}) in a NIR spectrum taken with the MOIRC
spectrograph on the Subaru telescope \citep{egami2008}.  The spectrum
of EGS\_b6, however, shows two emission lines at 27.7 and
31.1~\micron\ that we are not able to identify consistently with any
redshift.  EGS\_b6 is resolved to two objects 0\farcs7 apart in the
HST ACS image \citep{davis2007}, and an optical spectrum of this
system shows two galaxies at $z=1.02$ and $z=2.001$. We therefore omit
EGS\_b6 from the sample for further analysis.  The 24~\micron\ images
show several serendipitous objects in slits of all 12 targets, most of which
are too faint to permit redshift identification.
Only one source, EGS24a, have $F_{24~\micron}\sim 1$~mJy.  This object,
found in the slit of EGS24, shows the silicate absorption feature at
$z=2.12$ (Fig.~\ref{f:spec}).

\begin{figure}
\plotone{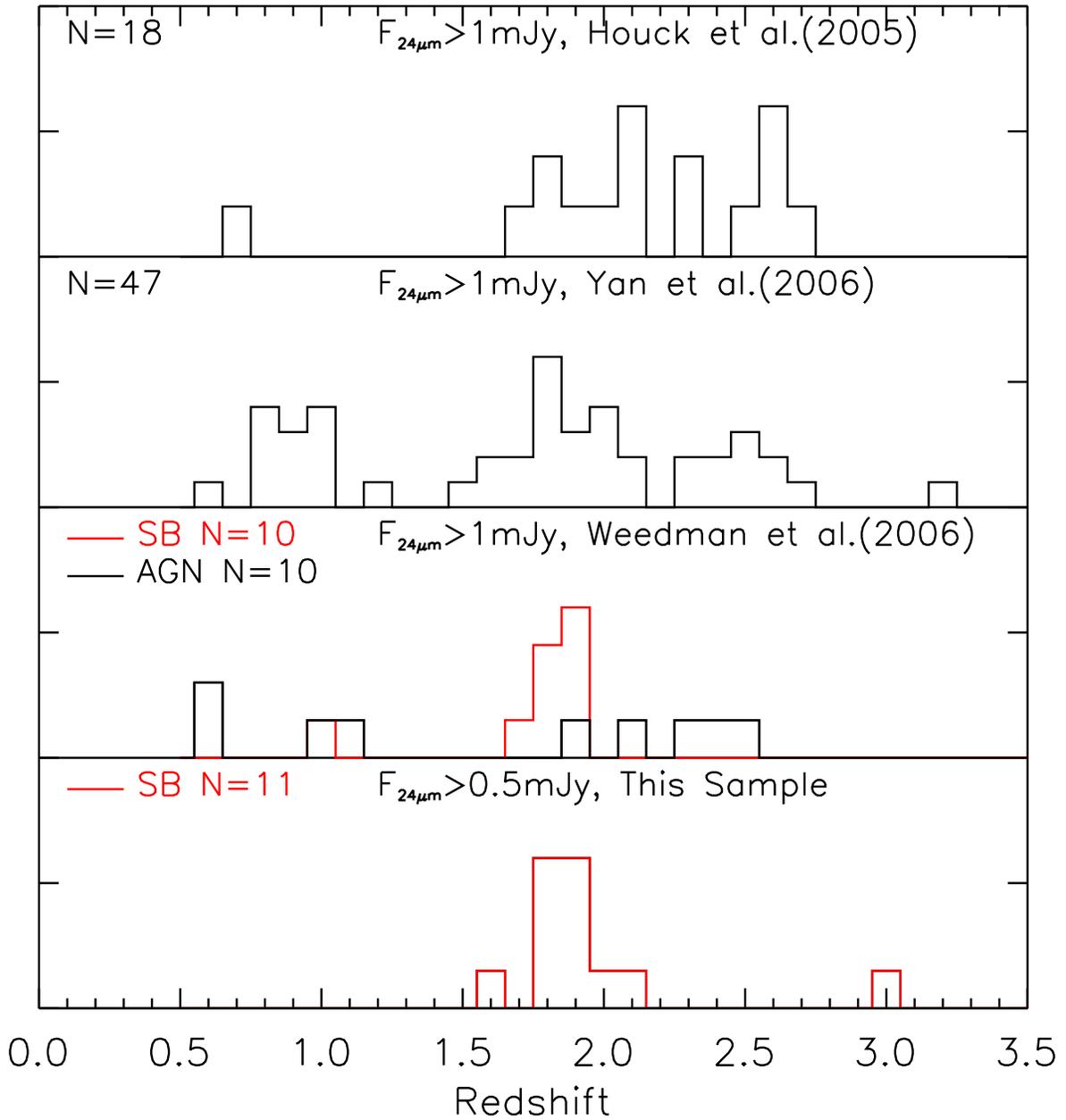}
\caption{Redshift distributions for spectroscopic samples in Table~1.
In the third panel, the red line shows the distribution for the
starburst (SB) sample and the black line for the AGN sample.
\label{f:zhist}}
\end{figure}

The redshift distribution of the sample (Fig.~\ref{f:zhist}) is very
similar to that of the starburst-dominated ULIRGs studied by
\citet{weedman2006}, even though our limiting flux density at
24\micron\ is a factor of two fainter than theirs. Recently
\citet{farrah2007} use the same criteria to select a larger sample in
the Lockman Hole region for the IRS observation and yield a very
similar redshift distribution.  The narrow distribution for
starburst-dominated ULIRGs is due to the selection of strong
7.7~\micron\ PAH emission by the MIPS 24~\micron\ band at $z\sim 1.9$.
The peak of the redshift distributions for \citet{weedman2006},
\citet{farrah2007}, and our sample is at this redshift, confirming the
selection effect.  On the other hand, luminous 24 \micron\ sources
with power-law SED have a much wider redshift range up to $z\sim 3$
\citep{houck2005, yan2005,weedman2006}, but they will not pass our
IRAC color criteria or the "bump" SED criterion in \citet{weedman2006}
and \citet{farrah2007}.


\subsection{PAH Emission Features in ULIRGs} 

\begin{figure}
\plotone{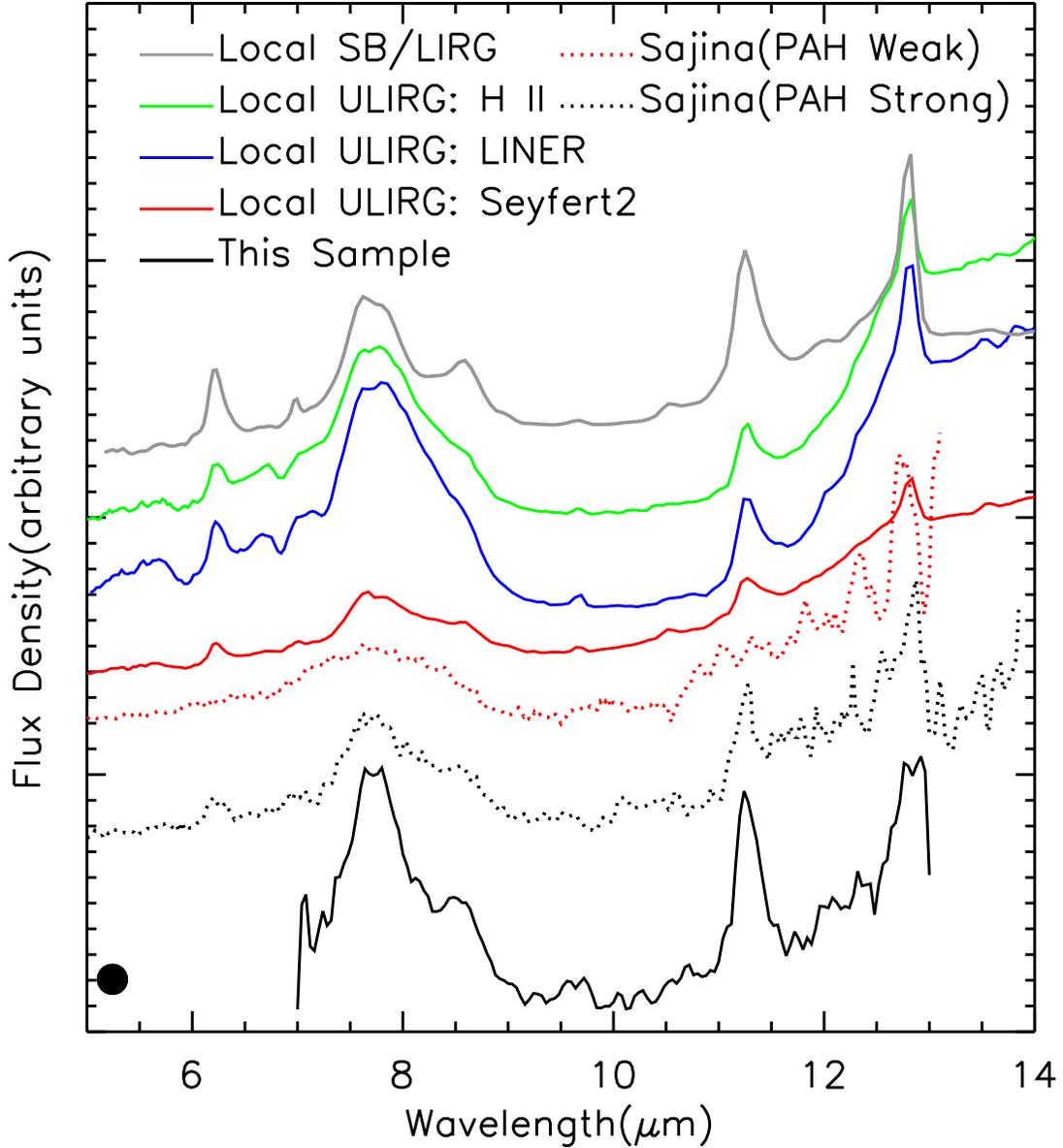}
\caption{Stacked spectrum for ULIRGs in the present sample (black
line). The short wavelength limit for the stacked spectrum is
7~\micron.  The large dot at rest wavelength 5.3~\micron\ represents
the stacked \AKARI\ 15~\micron\ flux density.  
The vertical scale is linear flux density per
unit frequency in arbitrary units.  Other lines show stacked spectra
of comparison samples: local starburst galaxies \citep[grey
line;][]{brandl2006}, local Seyfert-type ULIRGs (red line), local
LINER-type ULIRGs (blue line), and local \hii/starburst ULIRGs (green
line). The local ULIRG samples are from the \iras\
1~Jy sample \citep{kim1998} with IRS observations in the IRS GTO
program \citep[PID~105;][]{farrah2007,armus2007}.  Types were assigned
according to optical spectroscopy \citep{veilleux1995,
veilleux1999}. The average SED of the present sample is very similar to
those of local LINER and \hii-type ULIRGs, while average SEDs for objects in 
\citet{sajina2007a} are close to the local Seyfert
type ULIRGs with much higher continuum emission. 
\label{f:stack_sed}}
\end{figure}

The PAH features visible in the individual spectra of the sample
galaxies are even more prominent in the average spectrum for the
sample, as showed in Fig~\ref{f:stack_sed}, which also stacks local
starburst \citep{brandl2006} and ULIRG samples for comparison. The
local ULIRG sample is divided into Seyfert, LINER, and HII sub-samples
according to their optical spectral classification
\citep{veilleux1999}.  PAH emission features are found have different
feature profiles. \citet{Peeters2002} classified profiles of each PAH
emission feature, according to the peak wavelength, into 3 main
classes: Class A, B, and C. PAH emission features are known to have
more than one component in each feature.  For example, the
7.7~\micron\ PAH emission feature have two major components at 7.6 and
7.8~\micron : Class A is defined as 7.6\micron\ dominated PAH; Class B
is the 7.8\micron\ component dominated PAH; and Class C is red
component dominated with peak shifting beyond 7.8\micron.  The
7.7~\micron\ PAH in the local starburst spectrum appears to be more
consistent with class~A with the peak at wavelength shorter than
7.7~\micron. All local ULIRG spectra have a typical class~B PAH
profile, with a red wing extending beyond 8~\micron.  In \S3.1,
we already found that the starburst template fits each IRS spectrum of
our sample better than the ULIRG template. It is not surprising then
that the average 7.7~\micron\ PAH profile of our sample is more
similar to the average starburst spectrum, thus consistent with the
class~A.  Another significant difference is that our ULIRG sample has
an average $L_{11.3\micron}/L_{7.7\micron}$\footnote{$L_{11.3\micron}$
  and $L_{7.7\micron}$ are the 7.7 and 11.3\micron\ PAH emission
  luminosities defined as $L_{PAH}=4\pi d_L^2\int F_{PAH}(\nu)d\nu$}
ratio about twice as high as local ULIRGs but similar to local
starbursts \citep{brandl2006}.  We also plot the average spectra of
\citet{yan2005} in Figure~\ref{f:stack_sed} for comparison.  The average
spectrum for strong PAH objects in \citet{yan2005} is more similar to
the local Seyfert type ULIRG, implying a dominant AGN contribution in
the spectra of their sample.  We conclude from IRS stacking that the
7.7~\micron\ PAH profiles and $L_{11.3\micron}/L_{7.7\micron}$ ratios
for the present sample are more consistent with those of local
starburst galaxies rather than local ULIRGs.

PAH emission features are a tracer of star formation, one of the
energy sources powering ULIRGs \citep{genzel1998, rigopoulou1999,
  sajina2007a}.  In order to subtract the local continuum, we adopted
the method used by \citet{sajina2007a}, fitting the
$5<\lambda<15$~\micron\ spectrum into three components: the PAH
emission features, a power-law continuum, and the silicate absorption.
An iterative fit determined the continuum for each object.  The
initial input PAH template was from the NGC~7714 IRS spectrum after
subtracting its power-law continuum.  The silicate absorption profile
was from \citet{chiar2006} with central optical depth $\tau_{9.7}$ a
free parameter.  The 7.7 and 11.3~\micron\ PAH line luminosities and
equivalent widths for the local starburst sample \citep{brandl2006},
the local ULIRG sample \citep{armus2007}, and the present sample were
derived the same way.  \citet{brandl2006} used a different method to
derive the same parameters; their method would give lower 7.7\micron\ PAH flux
densities and luminosities. This is due to the complicated continuum
at $\sim 8$\micron.  Our 11.3\micron\ PAH flux densities are
consistent  with theirs. Table~3 gives the results.

\section{Multi-Wavelength Studies of ULIRGs at $z\sim 1.9 $}

\begin{figure}
\begin{flushleft}
\plotone{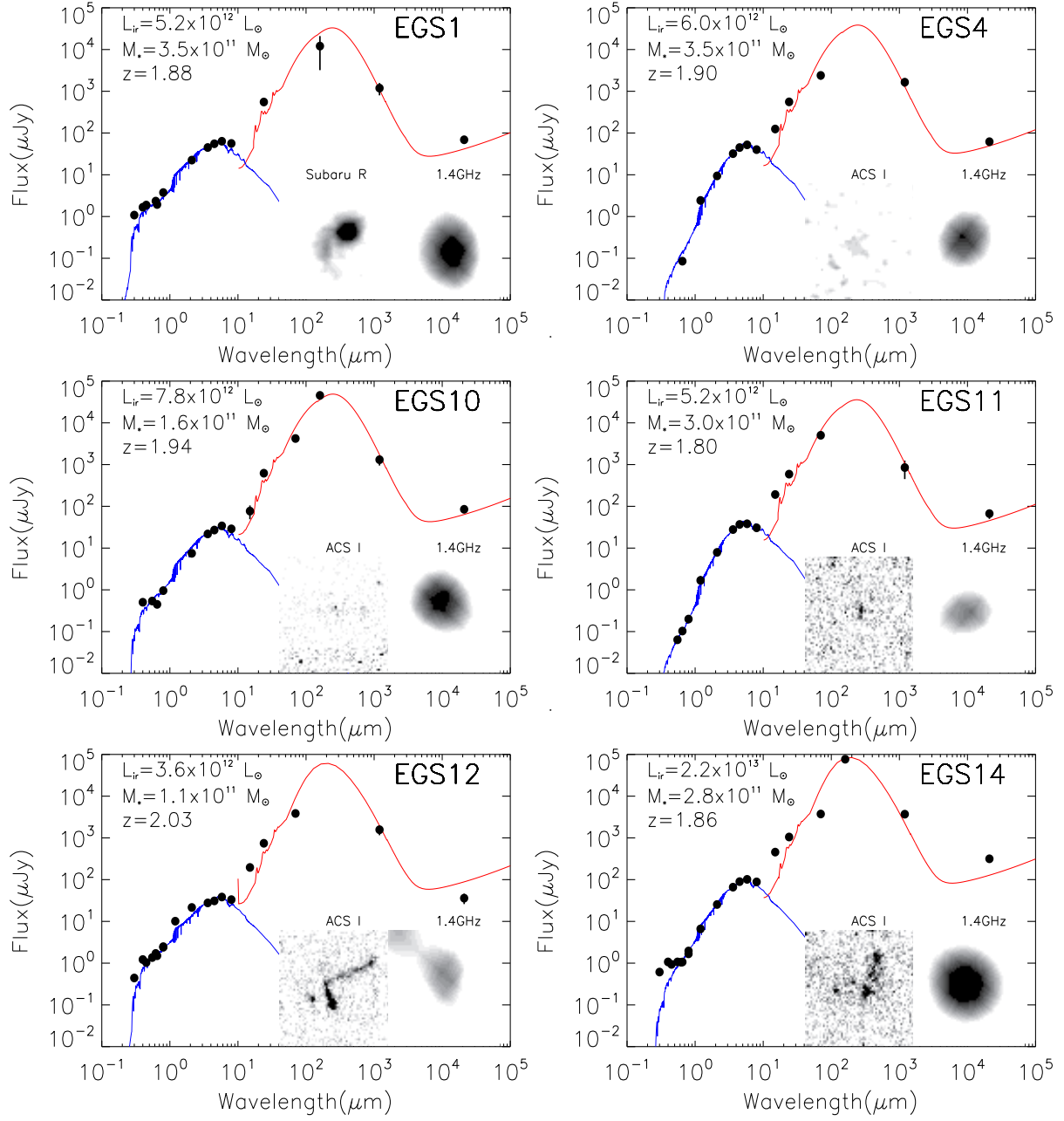}
\end{flushleft}
\caption{Spectral energy distributions and morphologies for sample
galaxies. Inset images are in negative grey scale and are 12\arcsec\
square.  The red images come from HST ACS data (filter F814W) if
available; otherwise Subaru $R$. Black dots represent photometric
data, and the blue line is the stellar population model (BC03) that
best fits each source.  The red lines are the CE01 dust templates
chosen to match the FIR luminosity of each source.
\label{f:sed}}
\end{figure}

AEGIS \citep{davis2007} and FIDEL \citep{dickinson2007} provide a rich
X-ray to radio data set to study the ULIRG SEDs. Objects in our sample
are measured at many key bands: all are detected in all four IRAC
bands \citep{barmby2008}, all but two by \AKARI\ at 15~\micron\
\citep{im2008}, and all at 1.4~GHz \citep{ivison2007}.  Most are also
detected at 70 and 160~\micron\ in the FIDEL survey
\citep{dickinson2007}. Only two objects, EGS14 and EGS\_b2, are
detected in the \chandra\ 200~ks X-ray imaging \citep{laird2008}.  The
flux densities in these key bands trace stellar mass, star formation
rate, and AGN activity. Objects in the present sample were also
observed with MAMBO on IRAM, and most were detected at 1.2~mm
\citep{younger2008}.  Table~4 gives the photometry, and the
UV-to-radio SEDs are shown in Figure~\ref{f:sed}.


The multi-wavelength photometry permits us to compare the present
sample with the sub-millimeter galaxy population.  There is a small
region covered by SCUBA in EGS by \citet{webb2003}, but no galaxies in
the present sample are in the SCUBA region. We fit FIR SEDs for the
sample and predict their 850~\micron\ flux densities $F_{850}$ to be
in the range $2.2 <F_{850}<8.4 ${\rm mJy} (Table~4). These values are
similar to the flux densities for SMGs at the same redshifts
\citep{chapman2006, pope06, pope08}.  The median $F_{850}$ for this
sample is 4.5 mJy, compared with the median $F_{850}$ of 5.5 mJy for
SMGs at $1.5<z<2.2$ found by \citet{chapman2006} and 7.5 mJy by
\citet{pope06,pope08}.  In more detail, 7 out 12 objects in the
present sample have $F_{850}$ fainter than 5 mJy, while the flux
densities for most SMGs in \citet{chapman2006} and
\citet{pope06,pope08} are brighter than 5 mJy.  We therefore argue
that this sample is part of a slightly faint SMG population.

Optical and radio morphologies of the galaxies provide important
information on their assembly histories.  {\it HST} ACS F814W imaging
\citep{lotz2008} covers the central 1\degr$\times$10\arcmin\ region of
the EGS.  EGS~1/4/b2 are outside the ACS coverage, but rough optical
morphologies are available from Subaru $R$-band images.  Optical images of
each object are presented with their SEDs in Figure~\ref{f:sed}.  Most
objects have irregular or clumpy morphologies in the rest-frame
$2000~\hbox{\AA}<\lambda<3000~\hbox{\AA}$ bands with a typical size of
1\farcs5, suggesting extended star formation in a region with a size
of about 13~kpc.  The 1.4~GHz radio imaging of EGS has a mean circular
beam width of $\sim$3\farcs8 FWHM \citep{ivison2007} and is unable to
resolve morphologies except in a few cases. EGS~23 and 24 show
elongated radio morphologies aligned with their optical extent,
indicating that the radio and rest-frame UV light are from the same
extended star formation regions in both cases. 

The spatial distribution of the stellar population is traced by the
rest-frame optical imaging.  \citet{windhorst2002},
\citet{papovich2005}, and \citet{conselice2005} have argued that
UV-dominated star-forming galaxies at high redshifts have similar
morphologies in the rest-frame UV and optical bands.  One outstanding
property of galaxies in the present sample is their extremely red
optical-NIR color. Seven objects in the sample have observed
$(R-K)_{\rm Vega}>5$, qualifying them as Extremely Red Objects (ERO).
EGS4 is the reddest with $(R-K)_{\rm Vega}=6.8$. Red colors like these
are common among distant ULIRGs; examples include ERO J164502+4626.4
(=[HR94]~10 or sometimes ``HR~10'') at $z=1.44$ \citep{elbaz2002} and
CFRS~14.1157 at $z=1.15$ \citep{lefloch2007}.  EROs are commonly seen
as counterparts to SMGs \citep{smail1999,frayer2004}. 
The red
optical-NIR colors, corresponding to rest $NUV - R$ for our sample,
indicate either dust extinction in these objects or high stellar mass. 
The stellar population modeling in the next paragraph suggests
objects in our sample have both heavy dust extinction and high
stellar masses. The heavy dust extinction does not seem to reconcile 
with objects being detected in the ACS 606W and 814W bands, which are 
the rest-frame 1800-2600\AA for galaxies at $z\sim 2$. The irregular and
clumpy morphologies in Figure~\ref{f:sed} 
highly non-uniform dust extinction in the objects in our sample.  
Only two objects are undetected in the deep ACS 814W
image, probably due to higher column density of dust in the compact stellar and gas distribution.

\subsection{Stellar Population and Mass in ULIRGs}

Stellar population modeling provides a way of determining physical
parameters from the observational data, but it is very difficult to
model stellar populations in ULIRGs.  \citet{tacconi2002} measured
dynamical masses for a sample of local ULIRGs with NIR spectroscopy
and found stellar masses in the range of $3\times10^{10} M_{\odot} <
M_{*} < 2.4\times10^{11} M_{\odot}$ with a mean of $1.1\times 10^{11}
M_{\odot}$. Their local sample has a mean absolute K-band magnitude of
$M_K=-25.8$ after adopting a dust correction of $A_K=0.7$ mag.  The
IRAC 8 \micron\ flux densities of this sample correspond to a similar
K-band magnitude range with $<M_K>=-25.7$ if the same dust correction
is used. This suggests a similar mass range for our sample, $M_* \sim
10^{11} \Msun$.

ULIRGs have a burst star formation history, very young stellar
populations, and non-uniform dust distribution, all of which can
introduce large uncertainties in modeling their stellar populations.
On the other hand, stellar masses are the most robust property against
variations in star formation history, metallicities, and extinction
law in modeling stellar population \citep{schreiber2004}.  We perform
a stellar population analysis on the present sample, mainly to measure
their stellar masses. We fit galaxy SEDs with \citet[][hereafter
BC03]{bc03} stellar population models with a Salpeter IMF and a
constant star formation rate.  Several groups \citep{shapley2001,
  dokkum2004, rigopoulou2006, kamson2007} have argued that a constant
star formation rate provides a reasonable description of stellar
population evolution for galaxies with ongoing star formation at high
redshifts, such as LBGs, Lyman-alpha emitters (LAEs), and DRGs.  The
stellar population age, dust reddening $E(B-V)$, stellar mass, and
derived star formation rate from the model fitting are listed in
Table~5, and the model SED fits are shown in
Figure~\ref{f:sed}. Objects in this sample have estimated stellar
masses with $M_{*}>10^{11}$~\Msun, similar to values found for local
ULIRGs \citep{tacconi2002}, DRGs, and {\it BzK} galaxies
\citep{labbe2005, daddi2007a}.  

\subsection{Total Infrared Luminosity and Star Formation Rate}
\label{s:lir}

\begin{figure}
\plotone{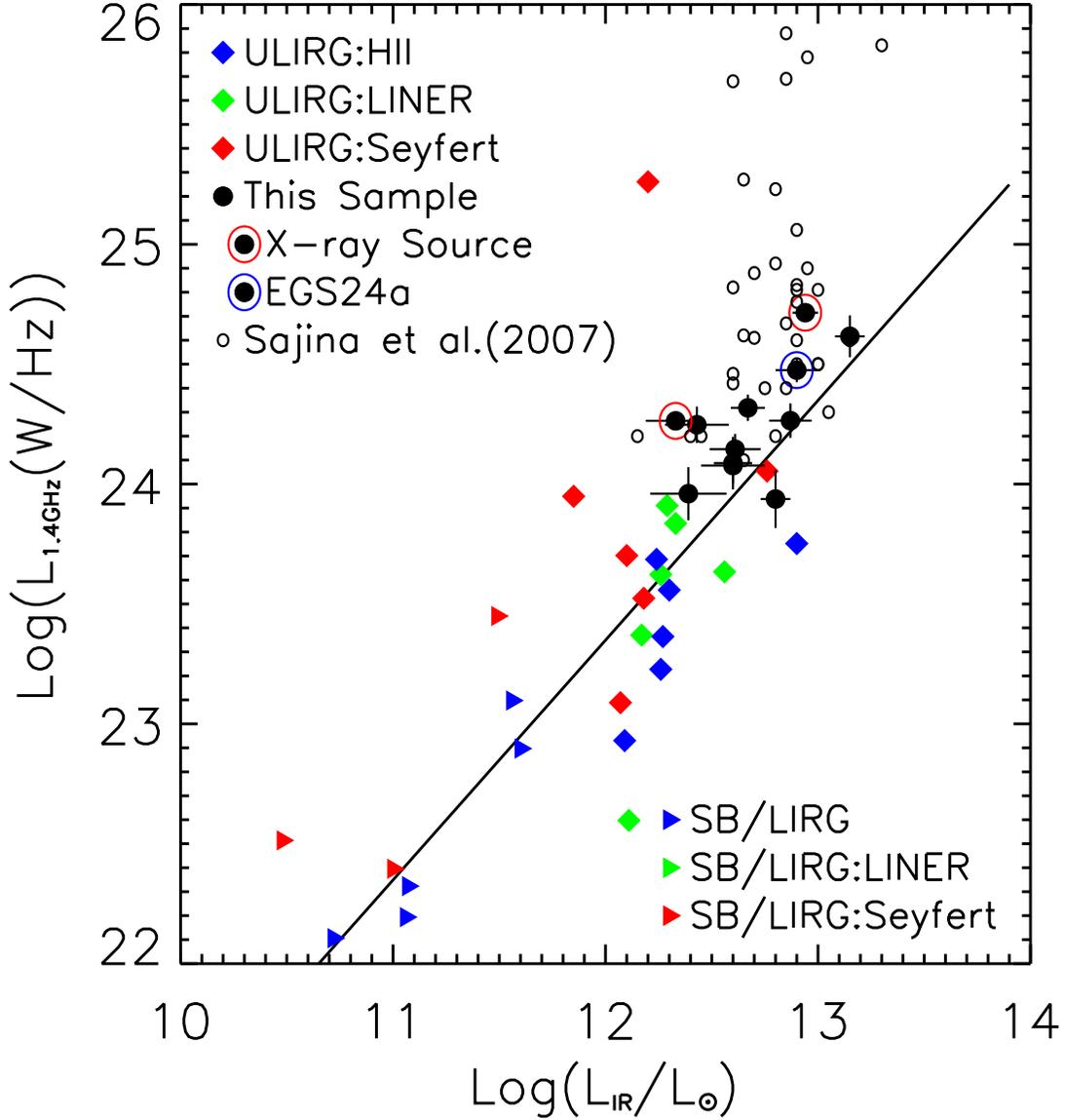}
\caption{Correlation between $L_{IR}$ and $L_{1.4GHz}$.  $L_{IR}$ was
  calculated by fitting SED templates (CE01) to the MIPS 70,
  160~\micron, and MAMBO 1.2~mm photometry. The two points with red
  circles are the X-ray sources EGS14 and EGS\_b2, and the one with a
  blue circle is the serendipitous object EGS24a. The sample is
  plotted together with local starburst galaxies and ULIRGs against
  the the local FIR-Radio relation (the thick line, Condon, 1992). The
  color coding: Seyferts in the local ULIRG and starburst samples are
  shown in red; LINERs in the local ULIRG and starburst samples are
  shown in green; and Starburst/HII-type ULIRGs in blue, the same as
  in Figure~\ref{f:stack_sed}.  This plot shows that the local
  starburst galaxies and ULIRGs, and objects in our sample are all
  consistent with the local FIR-Radio relation (Condon, 1992). Objects
  in \citet{sajina2007a} show strong radio excesses indicating AGNs in
  their sample.
\label{f:lir}}
\end{figure}

Two of the most-popular methods of estimating star formation rates of
local galaxies use total infrared luminosity $L_{IR}$ and radio
luminosity $L_{1.4GHz}$ \citep{condon1992, kennicutt1998}.  The
validity of these methods needs to be established at high
redshift. Most objects in the present sample are detected at
70~\micron, 160~\micron, and 1.2~mm, permitting a direct measurement
of total infrared luminosity $L_{IR}$ \citep{papovich2007}. In
practice, we derived $L_{IR}$ by fitting SED templates
\citep{chary2001} to the observed 70~\micron, 160~\micron, and 1.2~mm
flux densities (Fig.~\ref{f:sed}).  All galaxies in the sample have
$L_{IR}>10^{12}$~\Lsun\ (Table~4), qualifying them as ULIRGs. EGS14
and EGS21 have $L_{IR}>10^{13}$~\Lsun\ and are thus HyperLIRGs.  All
sample galaxies are also detected at 1.4~GHz \citep{ivison2007}.  We
will be able to verify: 1) whether $L_{1.4GHz}$ is correlated with
$L_{IR}$ for ULIRGs at $z\sim 2$; and 2) whether such a correlation at
high redshifts is consistent with the local one \citep{condon1992}.
Figure~\ref{f:lir} plots the radio luminosity $L_{1.4GHz}$ vs $L_{IR}$
for this sample and variety of local starburst and ULIRG samples.  
The FIR-radio ratio {\it q} for this sample\footnote{ $q={\rm
  log}(\frac{F_{FIR}}{3.75\time10^{12} W m^{-2}})-{\rm
  log}(\frac{F_{1.4GHz}}{W m^{-2} Hz^{-1}})$ defined by  
\citet{condon1992}} are in Table~4 with a mean $<q>=2.19\pm0.20$.
\citet{kovacs2006} measured $L_{IR}$ using 350\micron\, 850\micron\,
and 1.2mm flux densities and obtained a mean $<q>=2.07\pm0.3$ for SMGs
at $1<z<3$.  Both measurements yields {\it q} for ULIRGs at $z\sim 2$
close to, but smaller than the local value q=2.36. \citet{sajina2008}
showed more clearly a trend in their AGN dominated sample at $z\sim
2$: sources with strong PAH emission have {\it q} in $1.6<q<2.36$;
while all power-law sources have $q<1.6$. Normally, radio excess is
due to non-thermal emission from AGNs, but galaxy merging can also
enhance the non-thermal synchrotron radiation \citep{condon1992}.
Merging processes are evident in our sample. We will argue in
following paragraphs that AGNs activities may exist in most objects in
the sample.  In fact, two X-ray sources, EGS14 and EGS\_b2, and the
serendipitous power-law source EGS24a show higher radio excess (lower
{\it q}) than rest objects in the sample. Two scenarios can  be differentiated 
by their radio morphologies: AGNs are point sources and mergers, in most cases, are extended
sources.  Currently we cannot determine
which scenario is responsible to the radio excess due to lower resolution
of the 1.4gHz radio images (Figure~\ref{f:sed}).

Another measure used to estimate $L_{IR}$ for local galaxies is the
IRAC 8~\micron\ luminosity, $L_{8\micron}$, though there is
considerable debate about how reliable this method is.  $L_{8
  \micron}$ is defined as $L_{8\micron}=4\pi d_L^2 (\nu
F_{\nu})_{8\micron}$ where $F_{\nu}$ is the rest frame IRAC 8 \micron\
flux density \citep{huang2007b}.  $L_{8\micron}$ is found to be
correlated with $L_{IR}$ for local galaxies \citep{wu2005}. The MIPS
24 \micron\ band directly measures the rest IRAC 8 \micron\ flux
densities for our sample.  A galaxy's 8~\micron\ flux density actually
has two components (aside from starlight, which can be subtracted if
necessary): the 7.7~\micron\ PAH emission feature complex and a
featureless continuum, coming from an AGN or warm dust in the
interstellar medium.  There are several models for the IR emission
from galaxies, which convert $L_{8\micron}$ to $L_{IR}$
\citep[][hereafter CE01 and DH02]{chary2001,dh2002}.  Empirically,
\citet{wu2005} and \citet{bavouzet2008} found a correlation between $L_{8\micron}$ and both
$L_{1.4GHz}$ and $L_{IR}$ for star-forming galaxies.  
At the high luminosity end, local ULIRGs deviate from this
correlation with higher $L_{8\micron}$/$L_{IR}$ ratios, such a trend was also see by
\citet{rigby2008}.

\begin{figure}
\plotone{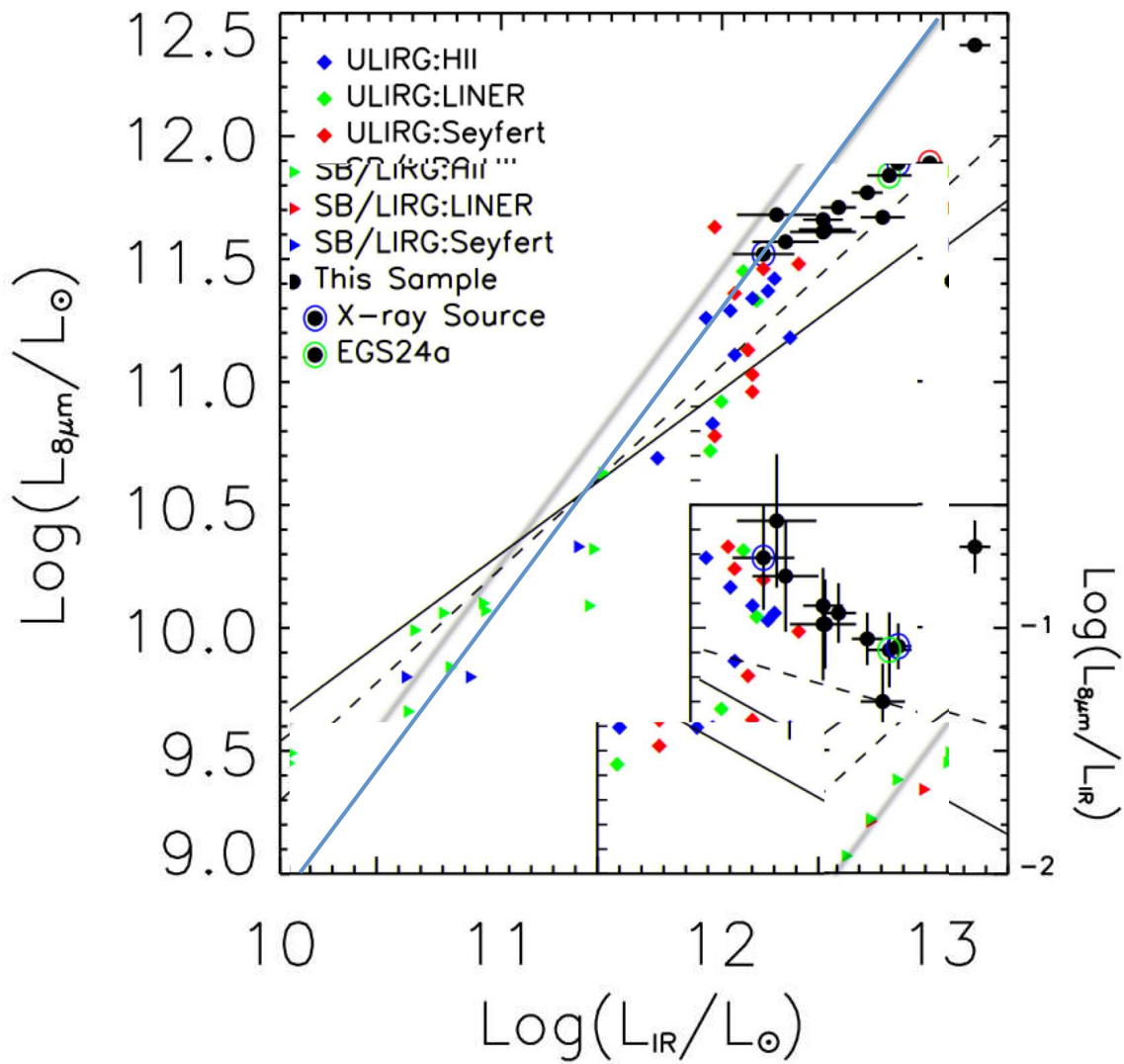}
\caption{Rest-frame 8~\micron\ luminosity, $L_{8~\micron}$ versus total infrared luminosity, $L_{IR}$. 
ULIRGs in this sample are shown as filled
black dots; the
two points with red circles are the X-ray sources EGS14 and EGS\_b2,
and the one with a blue circle is the serendipitous object
EGS24a; local starburst galaxies as triangles, and local ULIRGs as
diamonds.  Local starbursts and ULIRGs are color-coded based their
spectral classification \citep{veilleux1995,veilleux1999,
brandl2006}: red for Seyfert, green for LINER, and blue for
\hii/starburst. Two template models \citep{daddi2007a}  and one empirical \citep{bavouzet2008}
are also plotted: DH02 models as the dashed line, CE01 models as the solid line, and the empirical 
model as blue solid line.  The inset shows the same data for the high-luminosity galaxies
but plotted as the {\em ratio} of $L_{8~\micron}/L_{IR}$.
$L_{8~\micron}$ was measured for each galaxy by convolving its
spectrum with the redshifted bandpass of the IRAC 8~\micron\ filter.
In this plot, Rest-frame 8~\micron\ luminosities  for the present sample
are correlated with their total infrared luminosities,  but the observed $L_{8\micron}-L_{IR}$
relation is higher than both CE01 and DH02 model predictions. This implies that both models 
will over-estimate the $L_{IR}$ for the present sample based on their $L_{8\micron}$. 
\label{f:lum8}}
\end{figure}

Figure~\ref{f:lum8} shows a correlation between $L_{8\micron}$ and
$L_{IR}$ for all populations. However, the $L_{8\micron}-L_{IR}$
relation for objects in our sample and the local ULIRGs with high
$L_{IR}$ has a higher offset than that for the local starburst
galaxies and the model prediction \citep{chary2001,dh2002}. This
indicates that, for a given $L_{IR}$, $L_{8\micron}$ for objects in
our sample and some of local ULIRGs is higher than the model
prediction. Thus objects in our sample have an 8\micron\ excess
comparing with the CE01 and DH02 model prediction. The empirical $L_{8\micron}-L_{IR}$ 
relation of \citet{bavouzet2008}
derived with samples at various redshifts matches local starburst 
galaxies, but predicts much high $L_{8\micron}$ for ULIRGs and HyperLIRGs.
The $L_{8\micron}-L_{IR}$ relation for our sample permits to estimate $L_{IR}$
for same type of objects with only 24\micron\ flux densities.

\begin{figure}
\plotone{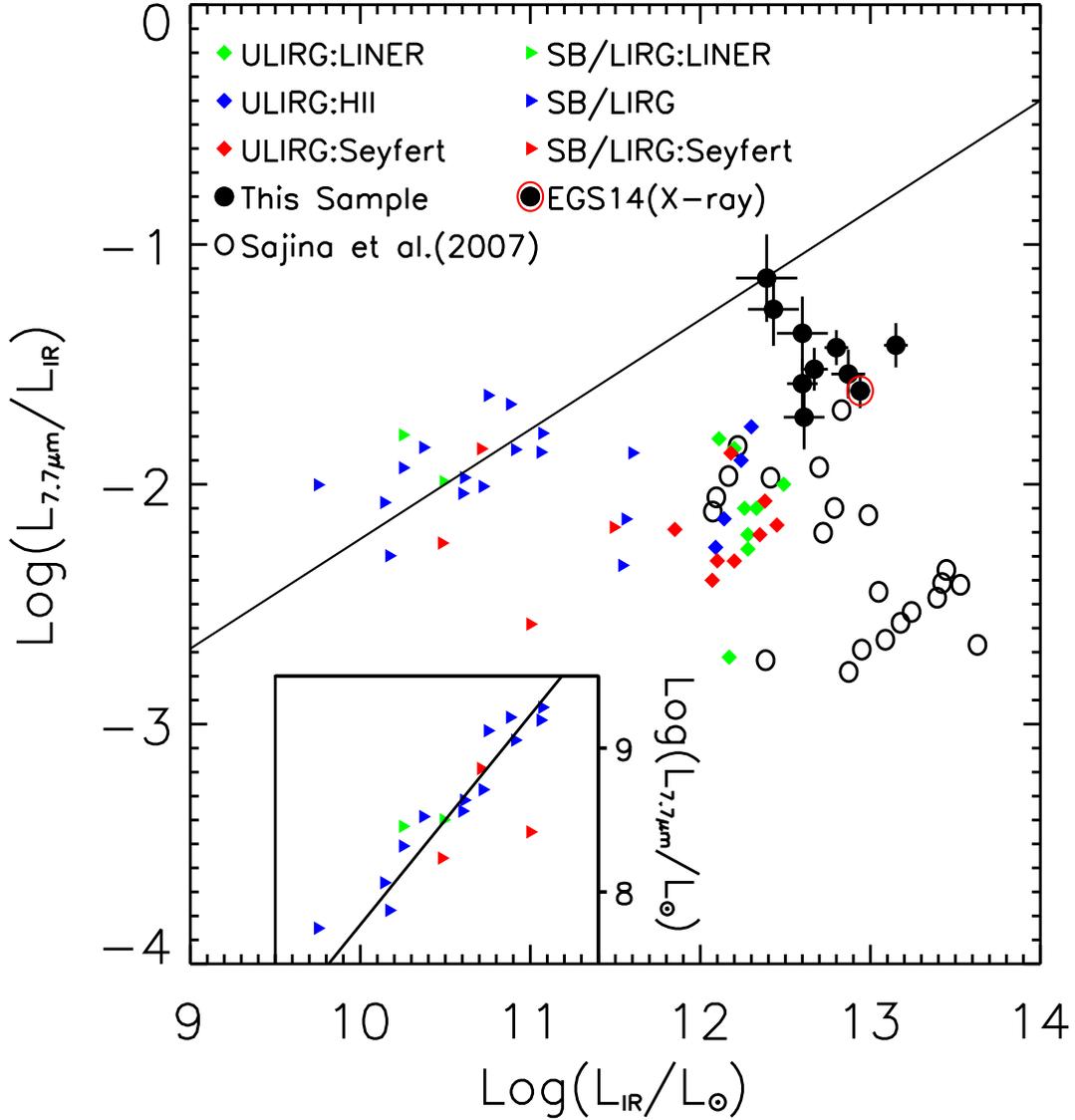}
\caption{The 7.7~\micron\ PAH to total infrared luminosity ratio versus total infrared luminosity.
The $L_{7.7}/L_{IR}$ ratio measure the star formation contribution in the total infrared luminosity
for objects in the sample. The present sample has the highest $L_{7.7}/L_{IR}$ ratio, indicating
that they are starburst dominated ULIRGs. The $L_{7.7}/L_{IR}$ 
ratio for the present sample is still compatible to the empirical relation 
of $L_{7.7}/L_{IR}\sim L_{IR}^{0.45}$ from the local starburst galaxies.
Objects in the present sample sample are shown as filled
black dots (the X-ray source EGS14 is a filled dots with red circle, the 7.7\micron\ PAH in the other X-ray sources EGS\_b2 
is not in our observation band, thus it is not in the diagram); local starburst galaxies as triangles, and local ULIRGs
as diamonds.  Local starbursts and ULIRGs are color-coded based on their
spectral classification \citep{veilleux1995,veilleux1999,
brandl2006}: red for Seyfert, green for LINER, and blue for
\hii/starburst. Open circles show ULIRGs and  HyperLIRGs from the \citet{yan2005}
sample with data from \citet{sajina2007a}.  The inserted plot shows a strong correlation between
$L_{7.7}$ and $L_{IR}$ for the local starburst galaxies. 
Thick lines show linear fits to the correlations for local
starburst galaxies as $L_{IR}\sim L_{7.7}^{0.69}$, which transfers to $L_{7.7}/L_{IR}\sim L_{IR}^{0.45}$
plotted with the thick line. 
\label{f:lum77}}
\end{figure}
\begin{figure}
\plotone{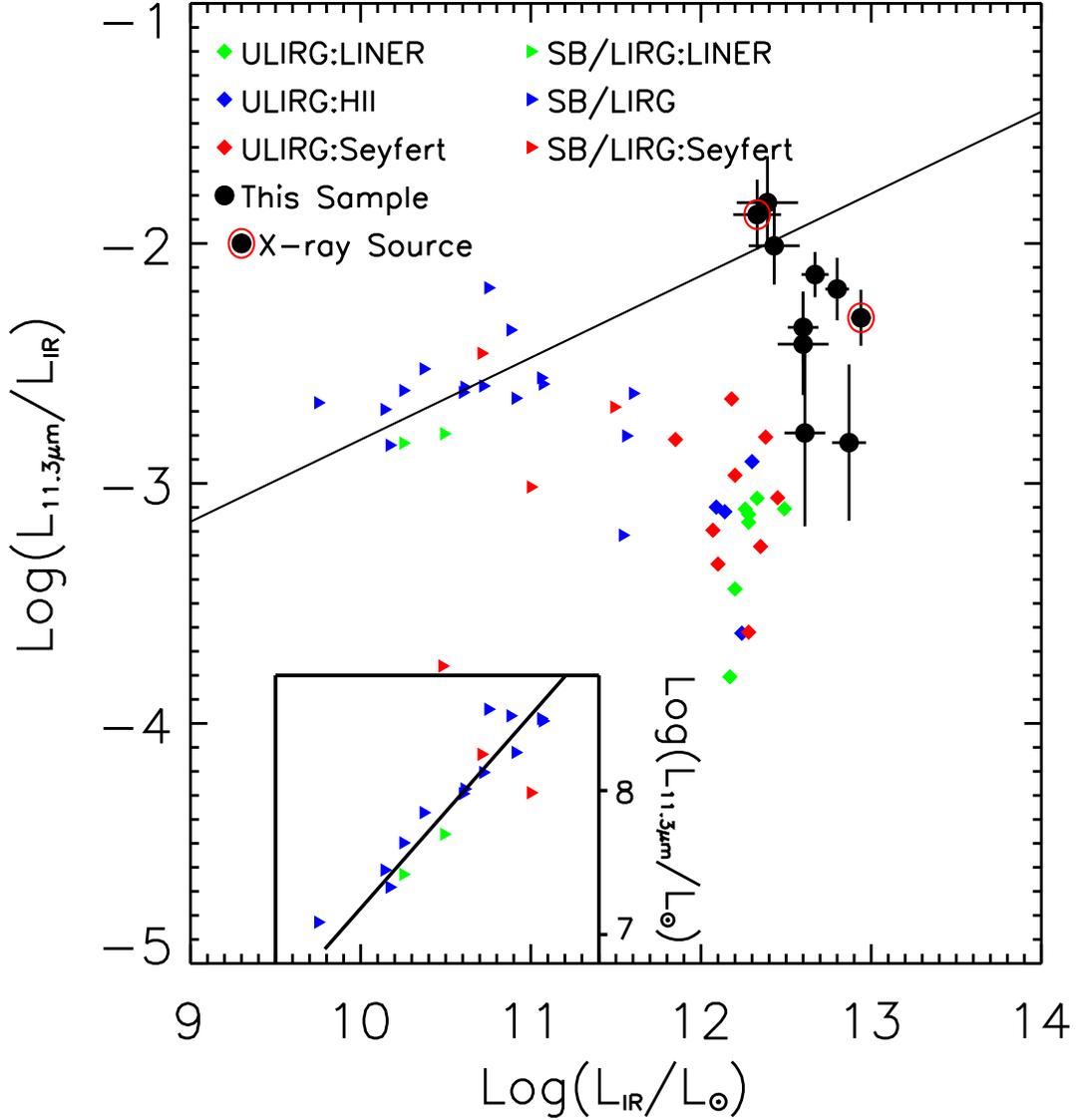}
\caption{The 11.3~\micron\ PAH to total infrared luminosity ratio versus total infrared luminosity.
The plot shows the same pattern as in Figure~\ref{f:lum77}.
Objects in the present sample sample are shown as filled
black dots; the two filled dots with red circles are two X-ray sources, EGS14 and EGS\_b2;
local starburst galaxies as triangles, and local ULIRGs
as diamonds.  Local starbursts and ULIRGs are color-coded based their
spectral classification \citep{veilleux1995,veilleux1999,
brandl2006}: red for Seyfert, green for LINER, and blue for
\hii/starburst.  The inserted plot shows a strong correlation between
$L_{11.3}$ and $L_{IR}$ for the local starburst galaxies. 
Thick lines show linear fits to the correlations for local
starburst galaxies as $L_{IR}\sim L_{11.3}^{0.75}$, which transfers to $L_{11.5}/L_{IR}\sim L_{IR}^{0.33}$
plotted with the thick line. 
\label{f:lum113}}
\end{figure}

Our IRS spectra can be used to separate the PAH from continuum in the
(rest) 8~\micron\ band, and each component's contribution to
$L_{8\micron}$ can be measured.  PAH luminosity is thought to be a
generally good tracer of star formation rate, but the
$L_{7.7\micron}/L_{IR}$ ratio is known to be luminosity-dependent,
decreasing at high luminosity
\citep{rigopoulou1999,desai2007,shi2007}.  Figure~\ref{f:lum77} shows
$L_{7.7\micron}/L_{IR}$ versus $L_{IR}$. In this diagram, each
population is well separated from the others.  The average
$L_{7.7\micron}/L_{IR}$ ratio for local ULIRGs is seen to be lower
than for local starburst galaxies. The HyperLIRGs in \citet{yan2005}
and \citet{sajina2007a} have the lowest $L_{7.7\micron}/L_{IR}$
ratio. In contrast, the present sample has the highest
$L_{PAH}/L_{IR}$ ratio, and the trend is the same for the
11.3~\micron\ PAH feature (Fig.~\ref{f:lum113}). Objects with such a
high PAH luminosity have neither been found locally nor in the MIPS
24\micron\ luminous sample at $z\sim 2$ \citep{yan2005, houck2005}.

Starburst galaxies were expected to have the highest $L_{PAH}/L_{IR}$,
and $L_{PAH}/L_{IR}$ was seen to decrease with increasing
$L_{IR}$. Our sample shows a new ULIRG population with much
higher PAH emissions at 7.7 and 11.3 \micron. We argue that the high
$L_{PAH}/L_{IR}$ ratio for our sample is generally compatible to
extrapolation from the $L_{PAH}/L_{IR}-L_{IR}$ relation for starburst
galaxies. 
Both $L_{7.7\micron}$ and $L_{11.3\micron}$ for local starburst
galaxies are strongly correlated with $L_{IR}$ in Fig.~\ref{f:lum77}
and Fig.~\ref{f:lum113}.  We fit both data sets and obtain the
following relations: $L_{IR}\propto (L_{7.7\micron})^{0.69}$ and
$L_{IR}\propto (L_{11.3\micron})^{0.75}$. Both relations convert to
$L_{7.7\micron}/L_{IR} \propto (L_{IR})^{0.45}$ and
$L_{11.3\micron}/L_{IR} \propto (L_{IR})^{0.33}$ respectively, as plotted
in Fig.~\ref{f:lum77} and Fig.~\ref{f:lum113}.  The $L_{PAH}$-$L_{IR}$
relation for local starbursts predicts a higher $L_{PAH}$-$L_{IR}$ ratio
in the $L_{IR}$ range for our sample. Our sample have high $L_{PAH}$-$L_{IR}$ ratio
close to the extrapolation comparing with other ULIRGs population, indicating
a starburst domination. The deficient PAH emission in our sample implies most likely
existence of AGN in our sample, though strong UV from intensive star forming region
can also destroy PAH.

The MIR spectral properties and $L_{PAH}/L_{IR}$ of our sample are
closer to local starburst galaxies, even though their $L_{IR}$ differs
by 2 orders of magnitude.  \citet{farrah2007} reached the same
conclusion by comparing silicate absorption strength for their sample
with those for local ULIRG and starburst galaxies, and they propose
six possible scenarios to explain the similarity between high redshift
ULIRGs and local starburst galaxies. Our multi-wavelength data set
provides further constrain on physical properties of our sample.  The
ACS I-band images (Figure~\ref{f:sed}) show multi-clumpy morphologies
extended to $>10$~Kpc size for most objects in our sample.  At $z\sim
2$, the observed I-band probes the rest NUV band, thus is sensitive to
star formation.  Local ULIRGs, however, have much more compact
morphologies in the GALEX NUV images \footnote{The GALEX UV morphologies for local 
starburst galaxies and local ULIRGs are from http://galex.stsci.edu/GalexView/.}. 
The extend morphologies of our
sample support both gas-rich merging and starburst geometry scenarios proposed
by \citet{farrah2007}.  In this scenario, the silicate dust column
density is reduced after star formation region is stretched to a large
scale during merging. The extended morphologies in rest NUV indicate
an extended star formation in our sample, thus extended distribution
of PAH emission. In such an extended distribution, more PAH 
can survive in strong UV radiation field from central AGN than those in a compact distribution.
This scenario thus explains the higher $L_{PAH}/L_{IR}$ in our sample than local ULIRGs.

\begin{figure}
\plotone{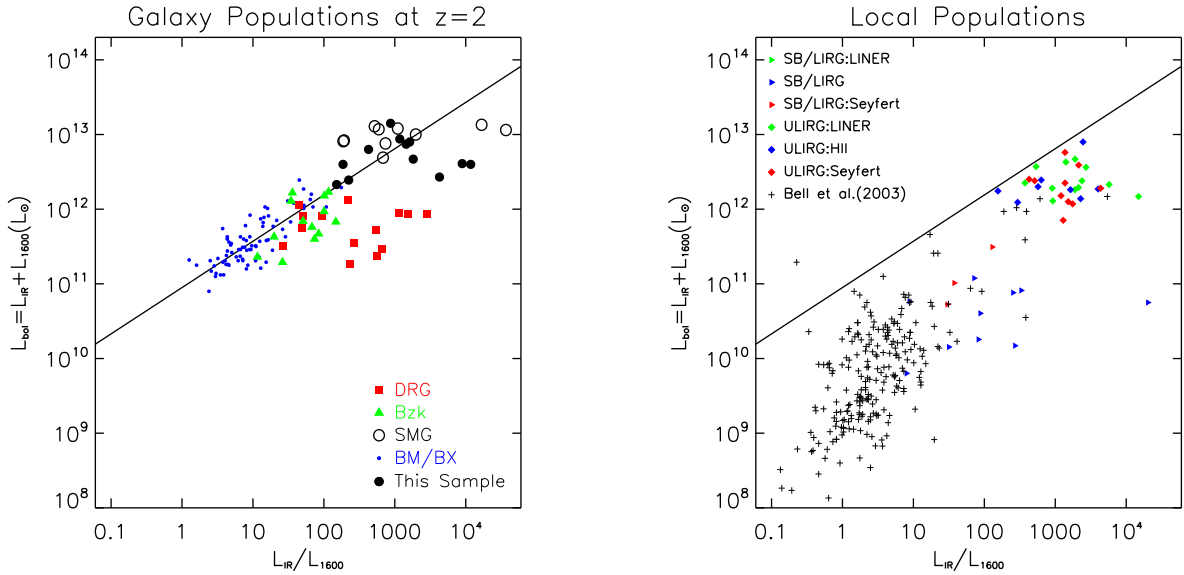}
\caption{The IR-to-UV luminosity ration for galaxies at $z\sim 2$ and
  $z\sim 0$. The left panel is for galaxies at $z\sim 2$. Our sample
  is plotted against galaxies at $z\sim 2$ selected in various
  bands. The solid lines in both panels are the
  $L_{IR}/L_{1600}$-$L_{IR}+L_{1600}$ relation for BM/BX sources, {\it
    BzK} galaxies, DRGs, and SMGs \citep{reddy2006}. The right panel
  is for local galaxies including normal galaxies \citep{bell2005},
  starburst galaxies \citep{brandl2006}, and ULIRGs. Most objects in
  our sample have the same relation as the rest of galaxies population
  at $z\sim 2$.  Three objects in our sample with extreme red colors,
  together with some DRGs in \citet{reddy2006}, are off the
  relation. They locate in the region where local ULIRGs are,
  indicating a compact dust distribution in those objects. See
  detailed discussion in the text.
\label{f:lir_1600}}
\end{figure}

Star forming galaxies at $z\sim 2$ are found to generally have much
less dust extinction than their local counterparts. \citet{reddy2006}
found that there is a correlation between $L_{IR}/L_{1600}$ and
$L_{1600}+L_{IR}$ for star forming galaxies at $z\sim 2$, where
$L_{1600}$ is the monolithic luminosity at 1600\AA.  This correlation
has a higher offset than the local relation, indicating less dust
extinction in the line of sight for galaxies at $z\sim 2$.  Most
objects in our sample lie upon the $L_{1600}/L_{IR}$-$L_{1600}+L_{IR}$
relation for galaxies $z\sim 2$
(Figure~\ref{f:lir_1600}). \citet{reddy2006} argued that dust
distribution and star formation region become more compact in local
galaxies.  We argue that lower surface density of dust density and
extended star formation region with high SFR permit to detect
both UV and PAH emission from most objects in ours sample.


 The star formation rate for a galaxy can be estimated from
its FIR and ultraviolet emission.  Specifically,
SFR is given as \citep{kennicutt1998, bell2005}
\begin{equation}
SFR/(\Msun~{\rm yr}^{-1}) =C \times
(L_{IR}+3.3L_{280})/\Lsun ,
\end{equation}
where $L_{280}$ is the monochromatic luminosity (uncorrected for dust
extinction) at rest frame 280~nm \citep{wolf2005}.  The constant {\it
  C} is $1.8\times 10^{-10}$ for the Salpeter IMF
\citep{kennicutt1998}, and $9.8 \times 10^{-11}$ for the Kroupa IMF
\citep{bell2005}. In the following text, we will adopt the Salpeter
IMF for all $L_{IR}$-{\rm SFR} conversion in this paper.  {\rm SFR}
will reduce by a factor of $\sim$2 if we switch to the Kroupa IMF
\citep{bell2005}.  The 280~nm band shifts to the observed $I$ band at
$z \approx 2$.  $L_{280}$ was calculated from the ACS F814W magnitude
if available or otherwise the CFHT $I$ magnitude.  All objects in our
sample have $L_{280}$ in the range $5\times10^8<L_{280}<3\times10^{10}
\Lsun$, less than 1\% of their $L_{IR}$.  The star formation rate seen
at rest-frame 280~nm is at most 20~\Msun~yr$^{-1}$, and most UV light
produced by newborn stars is absorbed by dust and re-emitted in the
infrared. Thus we omit the $L_{280}$ contribution in our SFR
calculation.

Total infrared luminosity, $L_{IR}$, of ULIRGs may be partly powered
by AGNs \citep{nardini2008}, thus using $L_{IR}$ may over-estimate
their SFR. The PAH emission only traces star formation, and is free of
AGN contamination.  We calculate {\rm SFR} for our sample with
their $L_{PAH}$ using the $L_{PAH}-{\rm SFR}$ relation, 
established from local starburst galaxies shown in
Figure~\ref{f:lum77} and Figure~\ref{f:lum113}.  Results are given
in Table 5. Star formation rates for our sample converted from
$L_{IR}$ using Equation~3 are much higher, with an average ${\rm SFR}
\sim 1000$~\Msun~yr$^{-1}$. $L_{7.7\micron}$ and $L_{11.3\micron}$
(Table~5) give smaller, star formation rates, in the range $150<{\rm
  SFR}<600$~\Msun~yr$^{-1}$ for most objects, that are quite
consistent with the stellar population modeling results.  The
discrepancy between both star formation estimations may be due to:
1. part of star formation occurs in region with no PAH, thus
$L_{PAH}$ underestimates the SFR; 2. $L_{IR}$ contains AGN
contribution, thus over-estimate the SFR.  It is very possible that
both can happen in one object simultaneously, namely its AGN destroys
PAH in surrounding area where star formation occurs. This will further
increase the discrepancy, so the real SFR should be in between both
estimations.

\begin{figure}
\plotone{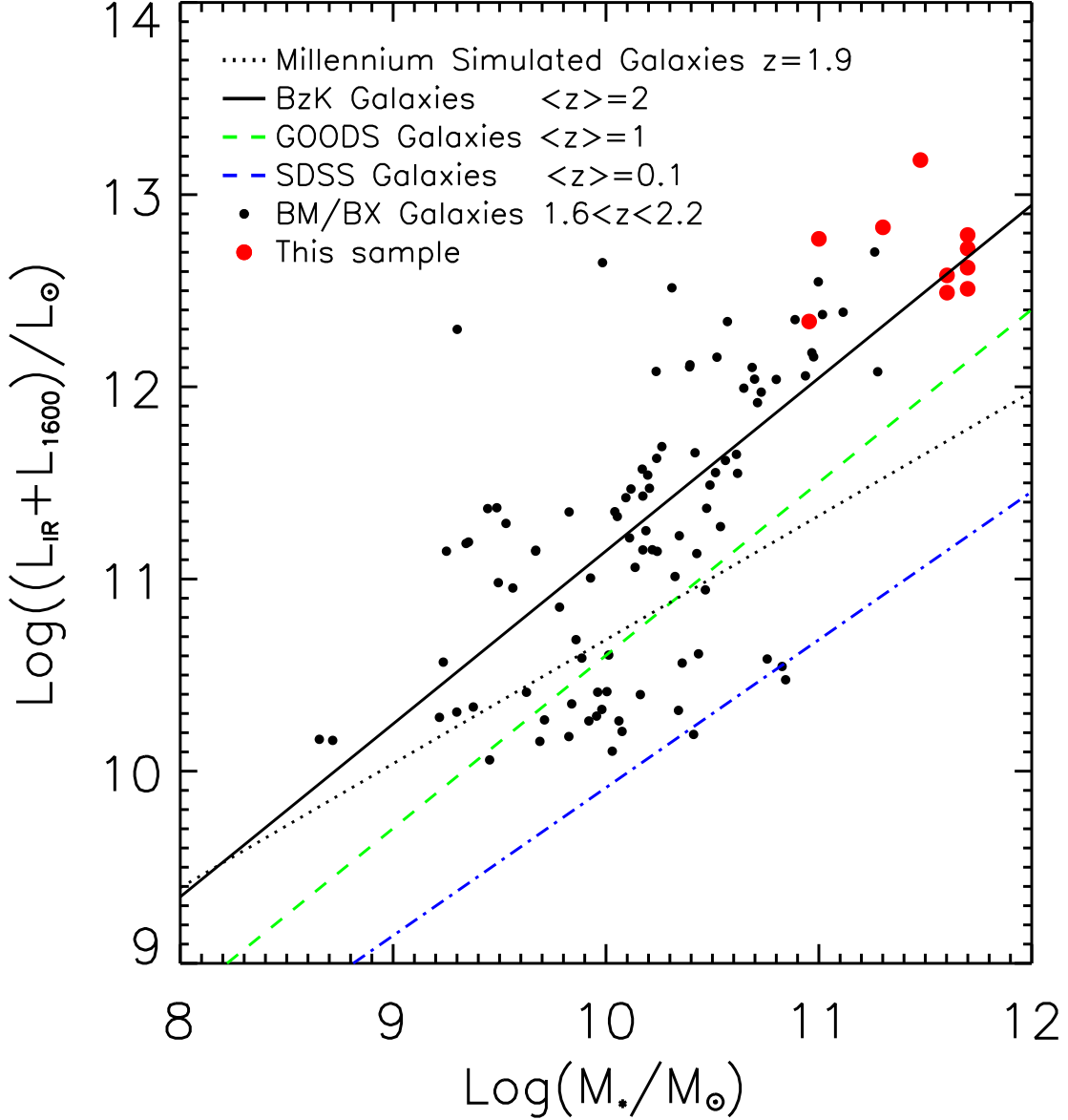}
\caption{The SFR-stellar mass relation for galaxies at z=0, 1, and 2 suggests the "downsizing" scenario
for galaxy formation.  The mean $L_{IR}$-$M_*$ relation for {\it BzK} at $z\sim 2$ is from \citet{daddi2007a},
the relations for GOODS galaxies at $z\sim 1$ and SDSS galaxies at $z\sim 0.1$ are from
\citet{elbaz2007}.
Again, objects in our sample are consistent with the relation for BM/BX sources and {\it BzK},
but at high mass end. Both simulated galaxy population models 
of \citet{kitzbichler2007} and 
\citet{cattaneo2008} predicts much lower star formation rate for galaxies with a given stellar mass. 
\label{f:lms}}
\end{figure}

Our sample have both high star formation rate and high stellar mass,
supporting the galaxy formation in the "downsizing" mode. The star
formation rates and stellar masses for our sample are consistent with
the SFR-stellar mass relation obtained from {\it BzK} galaxies at
$z\sim 2$ (Figure~\ref{f:lms}).  \citet{daddi2007a} showed that
simulated galaxy populations taken from the Milllennium Simulation
lightcones of \citet{kitzbichler2007} and \citet{cattaneo2008} failed
to re-produce the SFR-star mass relation at $z=2$, thus
underestimate number of ULIRGs at $z\sim 2$.

It has been long anticipated that ULIRGs have a dominant contribution
to the total infrared luminosity density, thus star formation rate
density, at $z\sim 2$ \citep{lefloch2005}.  We use the $V_{max}$
method to calculate the total infrared luminosity density for our
sample to be $2.6\times 10^7 L_{\odot}/Mpc^3$.  The sample of
\citet{farrah2007} with the same limiting flux yields a density of
$8.8\times10^6 L_{\odot}/Mpc^3$. We argue that the difference is due
to the cosmic variance, because these objects are massive galaxies and
thus have a much stronger spatial correlation.  Both densities are
lower than ULIRG $L_{IR}$ density at $z\sim 1$, $\sim 10^8
L_{\odot}/Mpc^3$ for all ULIRGs \citep{lefloch2005}. Most objects in
our sample and those of \citet{farrah2007} have $L_{IR}> 5\times
10^{12} L_{\odot}$, the
major contribution to the $L_{IR}$ density at $z\sim 2$ comes
from ULIRGs with $10^{12}<L_{IR}<5\times 10^{12} L_{\odot}$ \citep{huang2009}.

\subsection{AGN in the $z\sim 1.9$ ULIRG sample}

One direct way of identifying an object as an AGN is to measure its
X-ray luminosity.  Two objects in the sample, EGS14 and EGS\_b2, are
in the main AEGIS-X catalog from the \chandra\ 200~ks images
\citep{nandra2007, laird2008}.  Their X-ray fluxes $F_{0.5-10~keV}$
are $1.2\times10^{-15}$ and $6.4\times10^{-15}$~erg~cm$^{-2}$~s$^{-1}$
respectively. Calculated X-ray luminosities $L_X$ \citep{nandra2007,
  georgakakis2007} are $1.0\times10^{43}$~erg~s$^{-1}$ for EGS14 and
$9.4\times10^{43}$~erg~s$^{-1}$ for EGS\_b2.  Hardness ratios are 0.45
and -0.30, respectively.  Therefore EGS14 is a type~2 (obscured) AGN,
and EGS\_b2 is very close to a type~1 (unobscured) QSO according to
the X-ray luminosity and hardness ratios \citep{szokoly2004}. In
addition to EGS14 and EGS\_b2, EGS1 has a low-significance X-ray
counterpart.  At the location of this source there were 6.5 net soft
band counts (10 counts total with an estimated 3.5 count background).
This gives a Poisson probability of a false detection of
$3.5\times10^{-3}$.  The source was not detected in the hard band. If
the detection is real, EGS1 has $F_{0.5-2~keV}$~$ =
3.0\times10^{-16}$~erg~cm$^{-2}$~s$^{-1}$ with $L_{2-10~keV}$~$ =
1.1\times10^{43}$~erg~s$^{-1}$, thus is qualified to be an AGN.

The remaining 10 ULIRGs are not detected in the current Chandra observation.  Stacking in the soft band gives 19.5 counts above an
average background of 9.85, corresponding to
$F_{0.5-2~keV}=3.8\times10^{-17}$~erg~cm$^{-2}$~s$^{-1}$ or
$L_X=1.1\times10^{42}$~erg~s$^{-1}$ at 2$\sigma$ significance.
There was no detection in the hard band.  Even if EGS1 is added to
the stacking, nothing shows up in the hard band, but the soft band
detection significance rises to 3.2$\sigma$.  The mean flux is
$F_{0.5-2~keV}=4.7\times10^{-17}$~erg~cm$^{-2}$~s$^{-1}$ or
$L_X=1.3\times10^{42}$~erg~s$^{-1}$.  This average X-ray luminosity
represents either a very weak AGN or strong star formation. 
Using the relation of \citet{ranalli2003}, this average X-ray luminosity
corresponds to a star formation rate of 220 \Msun/yr, consistent with the SED and
PAH estimation. However, we argue that the stacked X-ray signal comes from central
point sources. These objects have very extended and elongated morphologies in the rest-frame
NUV band. If the X-ray photons are from these star formation regions, stacking would
not yield any signal unless they are aligned.

Emission in the rest 3--6~\micron\ wavelength range is another
indicator of AGN activity \citep{Carleton1987,shi2005, alonso2006,
hines2006, jiang2006, shi2007}. The longer end of that range, which
has minimal stellar and PAH emission contamination, is ideal for
detecting what is nowadays thought to be hot dust emission closely
related to the AGN accretion disk. Luminosity at
these wavelengths ($L_{5}$) can be converted to $L_{IR}$ for QSOs
with the QSO SED templates \citep{elvis1994}. \AKARI\ 15~\micron\
photometry \citep{im2008} provides the best measurement of $L_{5}$
for our sample.  All galaxies within the \AKARI\ coverage
are detected except EGS26, for which the 3$\sigma$ limiting flux
density is $F_{15}<58$~$\mu$Jy \citep{im2008}.  The \AKARI\
15~\micron\ band is wide enough to include the 6.2~\micron\ PAH
feature for objects with $1.5<z<2.2$, but this feature is much weaker
than the 7.7~\micron\ feature.  Thus the \AKARI\ 15~\micron\ band is
a better measure of AGN emission than the MIPS 24~\micron\ band. 

\begin{figure}
\plotone{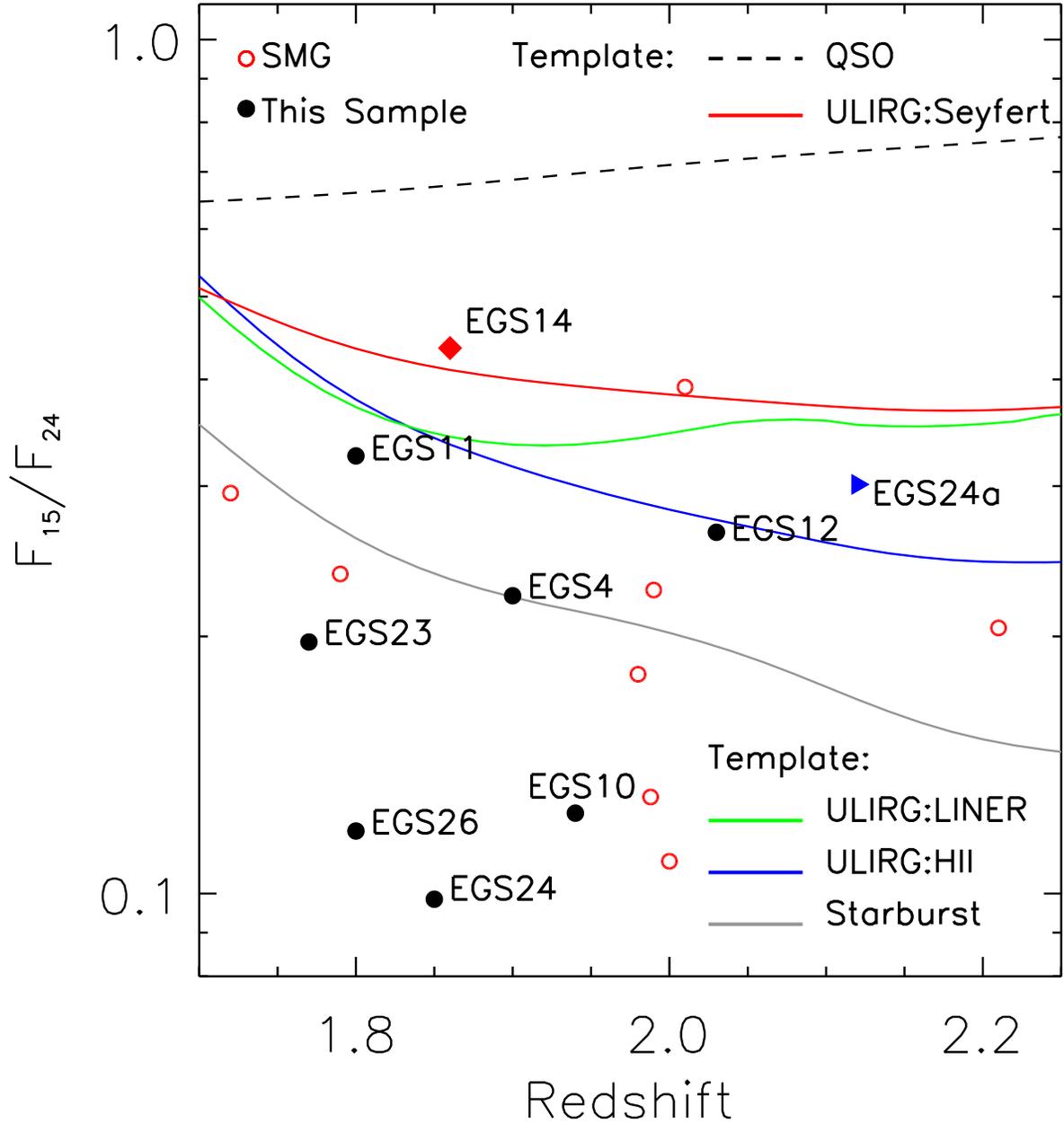}
\caption{$F(15~\micron)/F(24~\micron)$ versus redshifts for the ULIRG
sample (filled circles). The SMGs from \citet{pope08} are also plotted
(red open circles). This ratio should measure the continuum (hot dust) to PAH
ratio for objects in this redshift ranger and hence the AGN contribution to the mid-IR luminosity. 
Lines show the relations defined by local templates
(Fig.~\ref{f:stack_sed}): starburst galaxies (grey), Seyfert-type
ULIRGs (red), LINER-type ULIRGs (green), \hii/starburst ULIRGs
(blue), PG QSO (dashed line).  One X-ray source, EGS\_b2, is outside the \AKARI\
15~\micron\ coverage and not plotted.  The other X-ray source, EGS14
(red diamond), and the serendipitous object, EGS24a (blue triangle),
have colors consistent with AGN.  The 16~\micron\ flux densities 
for SMGs are measured from the IRS peak-up imaging. Since the IRS 16~\micron\ peak-up filter
profile is very similar to the AKARI 15~\micron\ filter profile, no correction is applied 
to the IRS 16~\micron\ flux densities for SMGs.
\label{f:ratio}}
\end{figure}

In fact, the $F_{15}/F_{24}$ ratio for our sample measures the
continuum-to-PAH ratio, and thus the AGN
fraction. Figure~\ref{f:ratio} shows this ratio versus redshift.  The
ratios for the two known AGNs with \AKARI\ coverage, EGS14 and EGS24a,
are very close to the expected values for Seyfert~2's.  EGS11 and
EGS12 are similar to expectations for \hii-type ULIRGs.  The flux
ratios for the remaining objects in our sample show even more PAH than
starbursts, indicating starburst domination in these objects. SMGs \citep{pope08} have
very similar $F_{15}/F_{24}$ ratios as objects in the present sample,
implying the same properties shared by both samples.  The SMGs also
show very strong PAH features in their IRS spectra.  This supports the
argument that most objects in the present sample are part of a SMG
population, and are starburst dominated ULIRGs.

A starburst dominated ULIRG can still have a deeply dust-obscured AGN.
Many current theoretical models \citep[e.g.,][]{mihos1994,
  mihos1996,dubinski1999, Cox2006, hopkins2006} suggest that such a
dust-obscured AGN can have a significant contribution to $L_{IR}$ of a
ULIRG. A study of local ULIRG IRS spectra show an average of 15\%
$L_{IR}$ are from central dust-obscured AGNs \citep{nardini2008}.
\citet{nardini2008} argued that ULIRG luminosity in $5<\lambda<6$ is
dominated by hot dust emission from AGNs.  Most objects in the present
sample are detected at 15\micron\, thus permit to measure their rest
5\micron\ luminosities, $L_{5 \micron}$, which trace AGN activity.
$L_{5 \micron}$ for the present sample is in range of $9.9< Log(L_{5
  \micron}/L_{\odot})<12.6$ (Table~4). Using $(L_{IR}/L_{5
  \micron})_{QSO}=22.8$ from the \citet{elvis1994} QSO SED, we
calculate that such a QSO contribution is about 14\% of $L_{IR}$ for
objects in our sample, consistent with those for local ULIRGs.

\section{Summary}

The results for the present sample combined with others in Table~1
show that high-redshift ULIRGs have a diverse range of properties, and
different selection criteria to pick out different populations.  The
combination of IRAC colors and MIPS 24~\micron\ flux used here selects
ULIRGs with strong 7.7~\micron\ PAH in a rather narrow redshift range
around $z\simeq 1.9$.  This sample shows a starburst dominated stage in
gas-rich merging powered ULIRGs at $z\sim 2$.  In this stage, intensive star
formation occurs in much extended region with a typical scale of $\sim 15$~Kpc
indicated by their ACS morphologies.  
Objects in this sample  have higher total infrared luminosities than local
ULIRGs, but the $L_{PAH}/L_{IR}$ ratios for the sample are higher than 
those of local ULIRGs.  We argue that the high
$L_{PAH}/L_{IR}$ ratio is due to the extended PAH distribution, which
is less affected by strong UV emission from central AGNs.
Most objects
follows the same $L_{IR}/L_{1600}$-$L_{bol}$  relation as that for
BM/BX, DRG and {\it BzK} galaxies, though they are at higher luminosity end. 

Stellar masses in this sample already exceed $10^{11}$~\Msun.  
Most stars must have formed prior to this stage. 
The SFR-stellar-mass relation for this sample is also consistent with that for 
the rest populations at $z\sim 2$, which is much higher than the theoretical model
prediction.
    
Only a few of the ULIRGs in our sample show direct evidence to have
AGNs with either high X-ray luminosities or hot dust emission in the
mid-infrared. Several pieces of evidence show weak AGNs existing in
this starburst dominated ULIRG sample: systematically higher
$L_{1.4GHz}/L_{IR}$ ratio than the local radio-FIR relation, and an
average X-ray emission of $L_X=1.3\times10^{42}$~erg~s$^{-1}$ from
point sources.  AGN contributes on average
15\% of total infrared luminosity for our sample.

This sample presents an early stage with very intensive star formation
but weak or heavily obscured AGNs.  ULIRGs in other samples at similar
redshift but with different selection methods \citep{yan2005,
  sajina2007a} have higher total infrared luminosities and lower PAH
luminosities, indicating increasing AGN and decreasing star formation
at higher $L_{IR}$.

\acknowledgements

This work is based in part on observations made with the Spitzer
Space Telescope, which is operated by the Jet Propulsion Laboratory,
California Institute of Technology under a contract with
NASA. Support for this work was provided by NASA through an award
issued by JPL/Caltech.   

Facilities: \facility{Spitzer}





\clearpage

\clearpage

\clearpage

\clearpage

\clearpage

\clearpage


\clearpage
\clearpage
\begin{flushleft}
\plotone{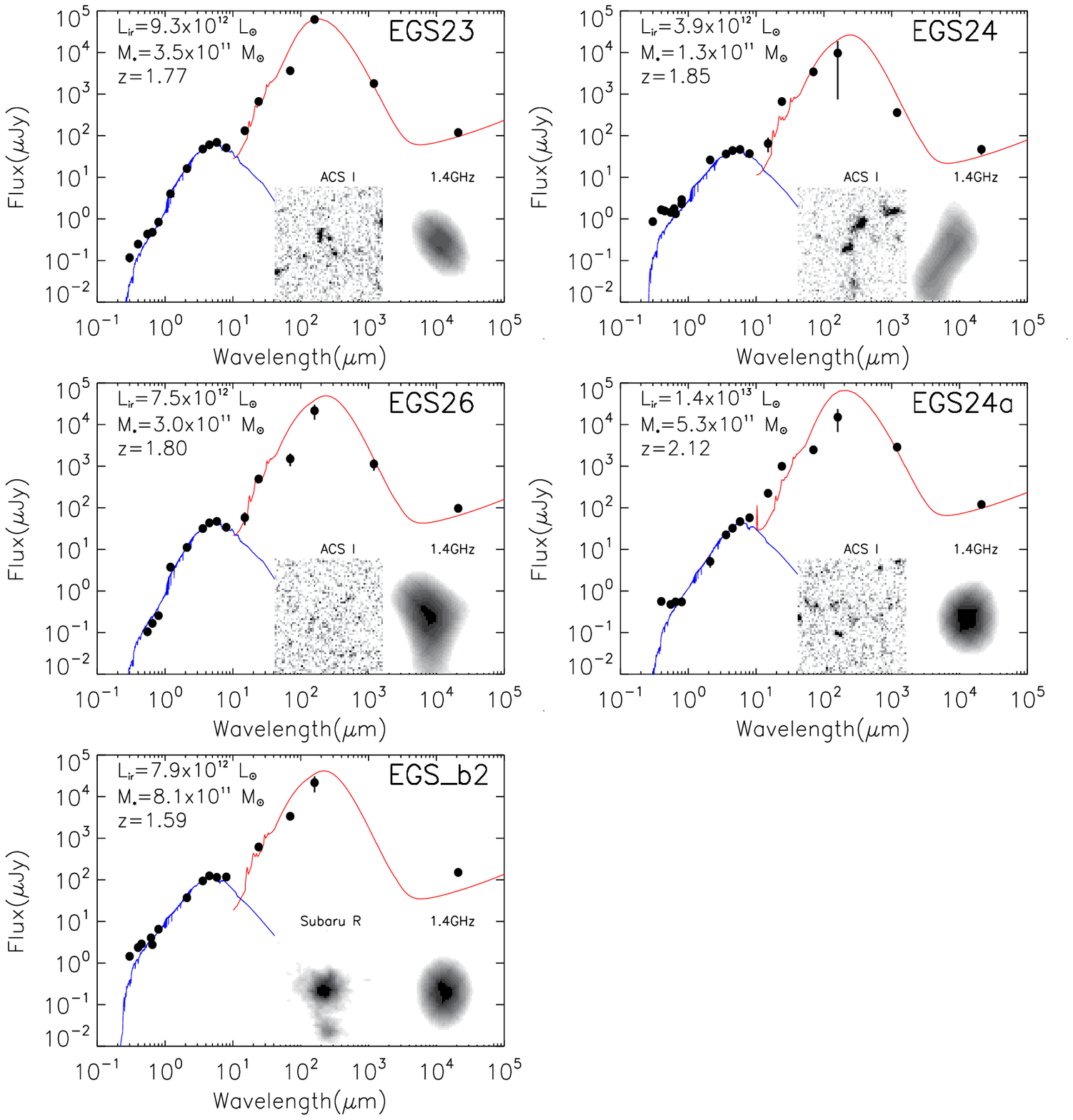}
\end{flushleft}
\clearpage

\clearpage

\clearpage

\clearpage

\clearpage

\clearpage

\clearpage

\clearpage



\begin{table*}[hbt]
{
\begin{center}
\centerline{\sc Table 1}
\centerline{\sc IRS Observation Parameters}
\begin{tabular}{lcccccc}
\hline\hline
Nickname\tablenotemark{a}
& EGSIRAC\tablenotemark{b}
& RA
& Dec
& $F(24\micron)$
& Cycles
& Exp time\\
&     &\multicolumn{2}{c}{J2000}& mJy & & s  \\
\hline
EGS1    & J142301.49+533222.4 & 14:23:01.50 & +53:32:22.6 & 0.55 & 10  & 7314\\
EGS4    & J142148.49+531534.5 & 14:21:48.49 & +53:15:34.5 & 0.56 & 10  & 7314\\
EGS10   & J141928.10+524342.1 & 14:19:28.09 & +52:43:42.2 & 0.62 & \08 & 5851\\
EGS11   & J141920.44+525037.7 & 14:19:17.44 & +52:49:21.5 & 0.59 & \08 & 5851\\
EGS12   & J141917.45+524921.5 & 14:19:20.45 & +52:50:37.9 & 0.74 & \05 & 3657\\
EGS14   & J141900.24+524948.3 & 14:19:00.27 & +52:49:48.1 & 1.05 & \03 & 2194\\
EGS23   & J141822.47+523937.7 & 14:18:22.48 & +52:39:37.9 & 0.67 & \07 & 5120\\
EGS24   & J141834.58+524505.9 & 14:18:34.55 & +52:45:06.3 & 0.66 & \07 & 5120\\
EGS24a\tablenotemark{c}  & J141836.77+524603.9 & 14:18:36.77 & +52:46:03.9 & 0.66 & \07 & 5120\\
EGS26   & J141746.22+523322.2 & 14:17:46.22 & +52:33:22.4 & 0.49 & 11  & 8045\\
EGS\_b2 & J142219.81+531950.3 & 14:22:19.80 & +53:19:50.4 & 0.62 & \08 & 5851\\
EGS\_b6 & J142102.68+530224.5 & 14:21:02.67 & +53:02:24.8 & 0.72 & \06 & 4388\\
\hline
\noalign{\hrule}
\noalign{\smallskip}
\end{tabular}
\tablenotetext{a}{Nicknames are the target names in the \s\ archive
and are used for convenience in this paper, but they are not official
names and should not be used as standalone source identifications.}
\tablenotetext{b}{Source name from \citet{barmby2008}.}
\tablenotetext{c}{Serendipitous source found in the slit while observing EGS24.}
\tablecomments{RA and Dec are the commanded telescope pointing
coordinates.  Telescope pointing was based on high accuracy peakup on
nearby 2MASS catalog objects with the blue peakup array.  Ramp
duration was 120~s for all objects.}
\end{center}
}
\label{tab1}

\end{table*}
\clearpage

\begin{table*}[hbt]
\begin{center}
\centerline{\sc Table 2}
\centerline{\sc IRS Sample Selection Criteria}
\begin{tabular}{lcl}
\hline\hline
Sample & 24~\micron\ flux density  & Color criteria \cr
\hline
\citet{houck2005} &  $>$0.75 mJy 
  & $\nu F_{\nu}(24\micron)/\nu F_{\nu}(I) > 60$\\
\citet{yan2005} & $>$0.9\0 mJy 
  & $\nu F_{\nu}(24\micron)/\nu F_{\nu}(I) > 10$ and\\ 
  &       & $\nu F_{\nu}(24\micron)/\nu F_{\nu}(8\micron) > 3.16$ \\
\citet{weedman2006}(AGN) & $>$1.0\0 mJy 
  & $F(\hbox{X-ray})\tablenotemark{a} \ga 10^{-15}$~erg~cm$^{-2}$~s$^{-1}$\\
\citet{weedman2006}(SB) & $>$1.0\0 mJy 
  & IRAC flux density peak at either 4.5 or 5.8\micron \\ 
This paper &  $>$0.5\0 mJy 
  & $0<[3.6]-[4.5]<0.4$ and \\
  &            &$-0.7<[3.6]-[8.0]<0.5$\\
\hline
\noalign{\hrule}
\noalign{\smallskip}
\end{tabular}
\end{center}
\label{tab2}
\tablenotetext{a}{{\it Chandra} 0.3--8~keV flux density}
\end{table*}

\clearpage
\begin{table*}[hbt]
{
\begin{center}
\centerline{\sc Table 3}
\centerline{\sc  PAH properties for the Sample}
\begin{tabular}{lcccccc}
\hline\hline 
Object & redshift$^a$ & redshift$^b$ & $\log L(7.7)$ &  $EW(7.7)$  &  $\log L(11.3)$  & $EW(11.3)$\cr
       &    &  & \Lsun       &  \micron    &    \Lsun          &  \micron \cr
\hline  
EGS1   & 1.95$\pm$0.03& 1.90$\pm$0.02&11.23$\pm$0.03  &  2.38$\pm$0.22        &      10.18$\pm$0.15    &    1.68$\pm$0.26   \cr   
EGS4   & 1.94$\pm$0.03& 1.88$\pm$0.02&10.89$\pm$0.06  &  0.57$\mp$0.07        &      ~~9.82$\pm$0.37    &   0.17$\pm$0.07    \cr   
EGS10  & 1.94$\pm$0.02& 1.94$\pm$0.01&11.33$\pm$0.02  &  2.39$\pm$0.12        &      10.04$\pm$0.31    &    0.26$\pm$0.10    \cr   
EGS11  & 1.80$\pm$0.02& 1.80$\pm$0.01&11.02$\pm$0.05  &  0.79$\pm$0.10        &      10.25$\pm$0.12    &    1.19$\pm$0.16    \cr   
EGS12  & 2.01$\pm$0.03& 2.02$\pm$0.03&11.37$\pm$0.02  &  1.46$\pm$0.08        &      10.61$\pm$0.11    &    1.28$\pm$0.55    \cr  
EGS14  & 1.87$\pm$0.06& 1.86$\pm$0.03&11.33$\pm$0.04  &  1.13$\pm$0.09        &      10.63$\pm$0.10    &    2.98$\pm$0.35    \cr   
EGS21  & 3.01$\pm$0.03& 3.00$\pm$0.03&11.73$\pm$0.06  &  1.59$\pm$0.10        &      \no               &    \no              \cr  
EGS23  & 1.77$\pm$0.02& 1.77$\pm$0.01&11.15$\pm$0.04  &  1.45$\pm$0.12        &      10.54$\pm$0.05    &    1.08$\pm$0.08    \cr   
EGS24  & 1.85$\pm$0.03& 1.85$\pm$0.01&11.25$\pm$0.03  &  2.24$\pm$0.18        &      10.56$\pm$0.07    &    0.36$\pm$0.08    \cr   
EGS26  & 1.77$\pm$0.03& 1.78$\pm$0.02&11.16$\pm$0.03  &  2.61$\pm$0.20        &      10.42$\pm$0.06    &    1.12$\pm$0.18    \cr   
EGS\_b2 & 1.59$\pm$0.01& 1.60$\pm$0.01& \no    &   \no        &      10.45$\pm$0.04    &   0.30$\pm$0.04    \cr
\hline
\noalign{\hrule}
\noalign{\smallskip}
\end{tabular}
\end{center}
\tablecomments{\no\ indicates the PAH feature lies outside the
observed spectral coverage. EGS24a shows silicate absorption and no PAH emission (Fig.3). }
}
\label{tab3}
\tablenotetext{a}{redshifts obtained with a ULIRG template.}
\tablenotetext{b}{redshifts obtained with a starburst template.} 
\end{table*}
  
\newpage

\clearpage
\begin{table*}[hbt]
\setlength{\tabcolsep}{0.02in}
{\scriptsize 
\begin{center}
\centerline{\sc Table 4}
\centerline{\sc IR/Radio flux and luminosity of the Sample}
\begin{tabular}{lccccccccccccc}
\hline\hline 
Object & $F(3.6~\micron)$ & $F(4.5~\micron)$ & $F(5.8~\micron)$ & $F(8.0~\micron)$ & $F(15~\micron)$ & $F(24~\micron)$ &  $F(70~\micron)$ & $F(160~\micron)$  & $F(850~\micron)$\tablenotemark{b} &  $F(1.1~mm)$
& $F(1.4~GHz)$ & $L_{IR}$\tablenotemark{c} &q \\  
 & $\mu$Jy & $\mu$Jy & $\mu$Jy & $\mu$Jy & $\mu$Jy & $\mu$Jy & mJy & mJy & mJy & mJy & mJy &\Lsun & \\
\hline  
EGS1   &  45.0$\pm$0.3&55.4$\pm$0.4&63.6$\pm$1.5&56.3$\pm$1.6&\no\tablenotemark{d}& 554$\pm$35& $<$1.5\0    &  12.1$\pm$8.9&3.3&1.86$\pm$0.50&0.069$\pm$0.010 &  12.72$\pm$0.15&2.25\\   
EGS4   &  32.1$\pm$0.3&44.9$\pm$0.4&52.1$\pm$1.5&40.1$\pm$1.5&125$\pm$24& 557$\pm$22&2.4$\pm$0.5  &  $<$21.0\0   &3.9&1.87$\pm$0.48&0.062$\pm$0.010 &  12.62$\pm$0.12&2.19\\   
EGS10  &  21.8$\pm$0.3&23.0$\pm$0.3&34.0$\pm$1.4&28.9$\pm$1.5&77$\pm$28& 623$\pm$35&4.2$\pm$0.7  &  45.5$\pm$8.7&5.2&1.65$\pm$0.69&0.085$\pm$0.014 &  12.83$\pm$0.10&2.33\\   
EGS11  &  27.8$\pm$0.3&36.9$\pm$0.4&38.2$\pm$1.4&30.6$\pm$1.5&192$\pm$31& 591$\pm$20&5.0$\pm$0.6  & $<$21.0\0    &3.3&0.85$\pm$0.44&0.067$\pm$0.017 &  12.58$\pm$0.09&2.24\\   
EGS12  &  27.7$\pm$0.3&31.0$\pm$0.3&38.3$\pm$1.4&33.2$\pm$1.5&196$\pm$25& 743$\pm$23&3.9$\pm$0.6  &  a           &5.4&1.58$\pm$0.47&0.036$\pm$0.010 &  12.77$\pm$0.07&2.59\\  
EGS14  &  66.1$\pm$0.2&89.6$\pm$0.4&101.7$\pm$1.5&88.4$\pm$1.6&457$\pm$39& 1053$\pm$41&3.8$\pm$0.6  &  76.7$\pm$9.6&6.4&4.54$\pm$0.68&0.316$\pm$0.023 &  13.18$\pm$0.06&1.95\\   
EGS21  &  39.5$\pm$0.3&45.3$\pm$0.4&50.7$\pm$1.5&35.0$\pm$1.5&59$\pm$14& 605$\pm$23&2.8$\pm$0.5  &  34.2$\pm$9.4&8.4&1.31$\pm$0.35&0.070$\pm$0.014 &  13.15$\pm$0.07&2.26\\  
EGS23  &  47.8$\pm$0.3&60.6$\pm$0.4&69.1$\pm$1.5&51.3$\pm$1.5&132$\pm$29& 665$\pm$18&3.7$\pm$0.4  &  62.4$\pm$8.7&4.5&1.81$\pm$0.40&0.119$\pm$0.015 &  12.79$\pm$0.08&2.08\\   
EGS24  &  36.4$\pm$0.3&44.0$\pm$0.4&46.8$\pm$1.4&37.1$\pm$1.5&65$\pm$25& 663$\pm$29&3.4$\pm$0.6  &  9.7$\pm$9.0 &2.7&1.49$\pm$0.74&0.047$\pm$0.012 &  12.51$\pm$0.18&2.16\\
EGS26  &  31.7$\pm$0.3&43.3$\pm$0.4&47.1$\pm$1.4&34.0$\pm$1.5&58$\pm$20& 492$\pm$16&1.5$\pm$0.5  &  21.6$\pm$8.4&4.5&1.14$\pm$0.36&0.097$\pm$0.017 &  12.49$\pm$0.15&2.15\\   
EGS24a &  22.3$\pm$0.3&32.3$\pm$0.3&46.7$\pm$1.5&575$\pm$1.6&223$\pm$36& 997$\pm$30&2.5$\pm$0.5  &  15.1$\pm$8.4&6.4&2.87$\pm$0.54&0.112$\pm$0.013 &  12.91$\pm$0.10&1.91\\  
EGS\_b2 & 94.0$\pm$0.2&124.8$\pm$0.4&115.0$\pm$1.5&117.1$\pm$1.6&\no            & 616$\pm$30&3.4$\pm$0.5  &  21.7$\pm$7.0&2.2&\no          &0.151$\pm$0.009 &  12.34$\pm$0.14&1.80\\
\hline
\noalign{\hrule}
\noalign{\smallskip}
\end{tabular}
\end{center}
}
\label{tab4}
\tablenotetext{a}{Confused}
\tablenotetext{b}{Prediction from the SED fitting.}
\tablenotetext{c}{FIR luminosity of the best-fit CE03 template.}
\tablenotetext{d}{not observed}
\end{table*}  
\newpage

\clearpage
\begin{table*}[hbt]
{
\begin{center}
\centerline{\sc Table 5}
\centerline{\sc Stellar Population Fitting Parameters}
\begin{tabular}{lcccc}
\hline\hline
Name
& Age
& $E(B-V)$
& $M_*$
& SFR\\
     & Gyr &   & $10^{11}$~\Msun & \Msun~yr$^{-1}$ \\
\hline
EGS1 & 1.9 & 0.3 & 5 & 240\\
EGS4 & 1.4 & 0.7 & 5 & 320\\
EGS10 & 1.1 & 0.4 & 2 & 182\\
EGS11 & 2.0 & 0.7 & 4 & 196\\
EGS12 & \00.29 & 0.4 & 1 & 480\\
EGS14\tablenotemark{a} & \00.26 & 0.6 & 3 & 1320\0\\
EGS23 & 1.1 & 0.6 & 5 & 400\\
EGS24 & \00.29 & 0.5 & 5 & 580\\
EGS26 & 1.8 & 0.6 & 4 & 220\\
EGS\_b2\tablenotemark{a} & \00.03 & 0.6 & \00.9 & 3800\0\\
\hline
\noalign{\hrule}
\noalign{\smallskip}
\end{tabular}
\end{center}
}
\label{tab5}
\tablenotetext{a}{EGS14 and EGS\_b2 are X-ray sources, and their SEDs
may be contaminated by AGNs.}
\tablecomments{EGS24a, the serendipitous object next to EGS24, was
not fit with any model because it has a power-law SED in the IRAC
bands.}
\end{table*}
\clearpage

\begin{table*}[hbt]
{
\begin{center}
\centerline{\sc Table 6}
\centerline{\sc Comparison of Star Formation Rates}
\begin{tabular}{lcccc}
\hline\hline
Name & SFR(BC03)\tablenotemark{a} & SFR(7.7)\tablenotemark{b} &
SFR(11.3)\tablenotemark{b} & SFR($L_{IR}$)\tablenotemark{c} \\ 
     & $M_{\odot}/yr$ &        &           &             \\
\hline
EGS1   &  240 &       386$\pm$18  &       263$\pm$\068 &         \0945$\pm$326 \\
EGS4   &  320 &       226$\pm$21 &        142$\pm$\090 &         \0750$\pm$207 \\
EGS10   & 182 &       452$\pm$14  &       207$\pm$110 &        1217$\pm$280 \\
EGS11   & 196 &       277$\pm$21  &       297$\pm$\061 &         \0 684$\pm$142  \\
EGS12   & 480 &       481$\pm$15  &       549$\pm$103 &        1060$\pm$171  \\
EGS14   & 1320\0 &    451$\pm$28  &         568$\pm$\097 &        2724$\pm$376  \\
EGS23   & 400  &      340$\pm$21  &         487$\pm$\042 &        1110$\pm$204 \\
EGS24   & 580  &      398$\pm$19  &         504$\pm$\060 &        \0582$\pm$241 \\
EGS26   &  220 &      346$\pm$16  &         390$\pm$\041 &        \0556$\pm$192 \\
EGS\_b2   & 3800\0 &   \no  &    417$\pm$\029 &         \0394$\pm$127 \\
\hline
\noalign{\hrule}
\noalign{\smallskip}
\end{tabular}
\end{center}
}
\label{tab6}
\tablenotetext{a} {Star formation rate calculated from stellar
population model (BC03) fitting} 
\tablenotetext{b} {Star Formation Rate calculated from PAH feature
luminosity.}  
\tablenotetext{c} {Star Formation Rate calculated from far infrared
luminosity $L_{IR}$} 
\end{table*}

\end{document}